\documentclass[10pt, twocolumn,twoside]{IEEEtran} %la original!!
\IEEEoverridecommandlockouts                              % This command is only needed if 
\usepackage{color}
\usepackage{algorithmicx}
\usepackage[ruled]{algorithm}
\usepackage{algpseudocode}
\usepackage{graphics} % for pdf, bitmapped graphics files
\usepackage{epsfig} % for postscript graphics files
\usepackage{subfigure}
\usepackage{ulem}
\usepackage{amsmath} % assumes amsmath package installed
\usepackage{amssymb}  % assumes amsmath package installed
\usepackage{tikz}

\newcommand{\R}{\mathbb{R}}
\newtheorem{theorem}{Theorem}[section]
\newtheorem{proposition}{Proposition}[section]
\newtheorem{definition}{Definition}[section]
\newtheorem{lemma}[definition]{Lemma}

\newtheorem{corollary}[definition]{Corollary}
\newtheorem{remarkth}[definition]{Remark}
\newenvironment{remark}{\begin{remarkth}\upshape}{\end{remarkth}}

\usepackage{amssymb}

\newcommand{\proa}{A^*G \mbox{$\;$}_{\tau^*} \kern-3pt\times_\alpha
G \mbox{$\;$}_\beta \kern-3pt\times_{\tau^*} A^*G}

\tikzstyle{vertex}=[circle,fill=black!20,minimum size=15pt,inner sep=0pt]
\tikzstyle{selected vertex} = [vertex, fill=red!24]
\tikzstyle{edge} = [draw,thick,-]
\tikzstyle{dedge} = [draw,thick,<->]
\tikzstyle{shadowdedge} = [draw, dotted,->]
\tikzstyle{weight} = [font=\small]
\tikzstyle{selected edge} = [draw,line width=3pt,-,red!50]
\tikzstyle{ignored edge} = [draw,line width=3pt,-,black!20]

\usepackage{nomencl}
\usepackage{ifthen}
\renewcommand{\nomgroup}[1]{%
\ifthenelse{\equal{#1}{C}}{\item[\textbf{Constants}]}{%
\ifthenelse{\equal{#1}{V}}{\item[\textbf{Variables}]}{%
\ifthenelse{\equal{#1}{S}}{\item[\textbf{sets}]}{}}}
}
\makenomenclature

\title{\LARGE \bf 
Forced Variational Integrators for the Formation Control of Multi-Agent Systems}

\author{Leonardo J. Colombo and H\'ector Garc\'ia de Marina

\thanks{L. Colombo (leo.colombo@icmat.es) is with Instituto de Ciencias Matem\'aticas (ICMAT), Calle Nicol\'as Cabrera 13-15, Cantoblanco, 28049, Madrid, Spain.  H. Garc\'ia de Marina (hgarciad@ucm.es) %is withUnmanned Aerial Systems Center, The Maersk McKinney Moller Institute,
is a Researcher in the Department of Computer Architecture and Automatic Control at the Faculty of Physics, Universidad Complutense de Madrid, 28040 Madrid, Spain}\thanks{L. Colombo and H. Garc\'ia de Marina were partially supported by I-Link Project (Ref: linkA20079) from CSIC. L. Colombo was partially supported by Ministerio de Economia,
Industria y Competitividad (MINEICO, Spain) under grant MTM2016-
76702-P; ''Severo Ochoa Programme for Centres of Excellence'' in R$\&$D
(SEV-2015-0554). The project that gave rise to these results received the
support of a fellowship from ``la Caixa'' Foundation (ID 100010434). The
fellowship code is LCF/BQ/PI19/11690016. The work of Hector Garcia de Marina is supported by the grant Atracci\'on de Talento with reference number 2019-T2/TIC-13503 from the Government of the Autonomous Community of Madrid.}}%
\begin{document}

\maketitle
\thispagestyle{empty}
\pagestyle{empty}

%%%%%%%%%%%%%%%%%%%%%%%%%%%%%%%%%%%%%%%%%%%%%%%%%%%%%%%%%%%%%%%%%%%%%%%%%%%%%%%%
\begin{abstract}
Formation control of autonomous agents can be seen as a physical system of individuals interacting with local potentials, and whose evolution can be described by a Lagrangian function. In this paper, we construct and implement forced variational integrators for the formation control of autonomous agents modeled by double integrators. In particular, we provide an accurate numerical integrator with a lower computational cost than traditional solutions. We find error estimations for the rate of the energy dissipated along with the agents' motion to achieve desired formations. Consequently, this permits to provide sufficient conditions on the simulation's time step for the convergence of discrete formation control systems such as the consensus problem in discrete systems. We present practical applications such as the rapid estimation of regions of attraction to desired shapes in distance-based formation control.
\end{abstract}
\section{Introduction}

Decentralized control strategies for multiple robotic systems have gained increased attention in the last decades in the control community. Distributed control algorithms for these systems can offer higher robustness and need for fewer resources per agent than centralized systems \cite{jorgebook}. In particular, formation control algorithms have emerged as powerful tools for the usage of multi-agent systems as surveyed by \cite{oh2015survey}. %The robots or agents are instructed to control geometrical relations, such as relative positions or distances, between their neighbors. As a result, the whole system displays a desired shape in the space.

Since the emergence of computational methods, fundamental properties such as accuracy, stability, convergence, and computational efficiency have been considered crucial for deciding the utility of a numerical algorithm. Geometric numerical integrators are concerned with numerical algorithms that preserve the system's fundamental physics by keeping the geometric properties of the dynamical system under study. The key idea of the structure-preserving approach is to treat the numerical method as a discrete dynamical system which approximates the continuous-time flow of the governing continuous-time differential equation, instead of focusing on the numerical approximation of a single trajectory. Such an approach allows a better understanding of the invariants and qualitative properties of the numerical method.

Using ideas from differential geometry, structure-preserving integrators have produced a variety of numerical methods for simulating systems described by ordinary differential equations preserving its qualitative features. In particular, from the engineering perspective, numerical methods based on discrete variational principles \cite{Lall1, marsden2001discrete} may exhibit superior numerical stability and structure-preserving capabilities. These methods can advance model-based design and analysis of networked control systems by preserving fidelity to the physical, continuous-time system, enabling, for instance, more accurate predictions of the energy transfer between agents as it is the case in formation control.

Variational integrators are numerical methods derived from the discretization of variational principles \cite{marsden2001discrete, hairer2006geometric}, \cite{Lall1}. These integrators retain some of the main geometric properties of the continuous systems, such as preservation of the manifold structure at each step of the algorithm, symplecticity, momentum conservation (as long as the symmetry survives the discretization procedure), and good (bounded) behavior of the energy associated to the system. This class of numerical methods has been applied to a wide range of problems in optimal control, constrained systems, power systems, nonholonomic systems, and systems on Lie groups. For more details we refer to \cite{ober2011discrete,leyendecker2010discrete,cortes2002geometric,kobilarov2011discrete}. 

Recently, the authors in \cite{sun2018conservation} studied conservation and associated decay laws in distance-based formation control of second order agents seen as a classical physical system. Following this approach inspired by classical systems, in this paper, we consider a more general class of systems by describing the dynamics of agents in the formation through a Lagrangian function or its associated Hamiltonian function, together with non-conservative (dissipative) forces. A similar mathematical description was recently proposed in \cite{colombo2018opt} and \cite{CoDiTAC} for the optimal control of multiple agents avoiding collision and in \cite{CD} for multi-agent motion feasibility systems with a Lagrangian dynamics. In this work, we study the construction and implementation of numerical methods for the formation control problem, where the desired formation is achieved by considering external (non-conservative) forces that dissipate the energy of the Lagrangian (conservative) system.

The implementation of variational integrators allows us to extend the study of (non-linear) formation control systems where it is not tractable to obtain non-conservative analytical results. For example, we can exploit variational integrators to study and characterize with accuracy the regions of attraction of the desired equilibrium or shape.

As a first result, the variational integrators can give sufficient conditions for the stability of formation control systems in discrete form, e.g., in numerical simulations with a fixed time step. We note that a particular case in formation control is the {\it{rendezvous}} of the agents, i.e., we have a (discrete-time) consensus problem \cite{zhucon10,wangcon11}. We can further employ the variational integrators for high accuracy numerical solutions without compromising the computational cost. In fact, a multi-agent system can consist of a significant number of agents and links where the bigger the number of initial conditions, the bigger the sensitivity for the agents' trajectories. For example, we have that desired shapes in the non-linear distance-based control are locally stable, and their analytic region of attraction is rather conservative, e.g., stability around a linearized system. Hence, the identification of larger regions of attraction needs to have accurate simulations of trajectories without dramatically increasing the computational cost with the number of agents.

In this paper, we introduce a mathematical framework based on tools of differential geometry to describe the formation control of multiple Lagrangian and Hamiltonian systems, and we construct a  geometric integrator based on the discretization of an extension of the Lagrange-d'Alembert principle for a single agent, in the spirit of \textit{forced variational integrators} \cite{marsden2001discrete}, \cite{ober2011discrete}. This is because in formation control the interaction between the agents can be described by conservative forces coming from local potentials such as \textit{elastic} ones. Such stored energy between neighboring agents is then dissipated by non-conservative forces in order to achieve the desired shape in the formation. This class of variational integrators has been recently studied in \cite{MdDSA}, \cite{izadi}, but not exploited for distributed control purposes. In particular, we construct and implement forced variational integrators for formation control of autonomous agents based on local potentials, and further, we provide an accurate numerical integrator with a lower computational cost than traditional solutions such as the ones obtained with a Runge Kutta method. We also find error estimations for the rate of the energy dissipated along the motion of the agents to achieve desired formations. This is done by defining a modification of the Hamiltonian vector field describing the dynamics of the continous-time system, and by studying backward error analysis for \textit{forced} variational integrators. One of the original contributions of this paper is the extension of the construction provided for unforced geometric integrators in \cite{hansen}. Such a non-trivial extension allows us to find bounds on the step-size of the proposed integration scheme for the rate of energy decay associated with a Hamiltonian function for the modified Hamiltonian vector field. Consequently, this permits to provide sufficient conditions for the convergence of discrete formation control systems.
%which is determined by the modified Hamiltonian vector field for the formation control of multi-agent systems
The remainder of the paper is organized as follows. In Section \ref{section2} we introduce variational integrators and the preliminaries definitions on the geometry and numerical aspects of Hamiltonian systems. In Section \ref{sec: LA}, we derive the dynamics for the formation control of multiple Lagrangian systems subject to external forces from Lagrange-d'Alembert principle. In Section \ref{sec: fc}, we construct forced variational integrators for the formation control of multi-agent systems derived by the discretization of the variational principle presented in Section \ref{sec: LA}. In Section \ref{sec: ham}, we introduce the Legendre transformation in both, continuous-time and discrete-time situations, to next construct the discrete Hamiltonian flow for formation control, which is used in Section \ref{sec: energy} to study the rate of dissipation  at each step of the algorithm. We show how to derive the discretized equations of motion and system's energy for generic formation controllers in Section \ref{sec: star}, and then we illustrate and compare the effectiveness of the proposed variational integration with numerical experiments. In the same section, we exploit the congervence guarantees to investigate regions of convergence beyond the conservative local values in distance-based formation control. Finally, we wrap up the presented work with some conclusions in Section \ref{sec: con}.

\section{Preliminaries}\label{section2}
\subsection{Discrete mechanics and variational integrators}
\label{sec: div} Let $Q$ be a $n$-dimensional differentiable
manifold with local coordinates $(q^A)$, $1\leq A\leq n$, the
configuration space of a mechanical system. Denote by $TQ$ its
tangent bundle, that is, if $T_{q}Q$ denotes the tangent space of $Q$ at the point $q$, then $\displaystyle{TQ:=\bigcup_{q\in Q}T_{q}Q}$, with induced local coordinates $(q^A, \dot{q}^A)$.  $T_{q}Q$ has a vector space structure, so we may consider its dual space, $T^{*}_{q}Q$ and define the cotangent bundle as $\displaystyle{T^{*}Q:=\bigcup_{q\in Q}T^{*}_{q}Q},$ with local coordinates $(q^A,p_A)$.

Given a Lagrangian function $L:TQ\rightarrow \R$, its Euler-Lagrange
equations are
\begin{equation}\label{qwer}
\frac{d}{dt}\left(\frac{\partial L}{\partial\dot
q^A}\right)-\frac{\partial L}{\partial q^A}=0, \quad 1\leq A\leq n.
\end{equation}

Equations \eqref{qwer} determine a system of $n$ second-order
differential equations. If we assume that the Lagrangian is regular,
i.e., the ${(n\times n)}$ matrix $\left(\frac{\partial^{2} L}{\partial \dot q^A\partial \dot q^B}\right)$, $1\leq A, B\leq n$, is non-singular, the local existence and uniqueness of solutions is guaranteed for any given initial condition by employing the implicit function Theorem.

A Hamiltonian function $H:T^{*}Q\to\mathbb{R}$ is described by the total energy of a mechanical system. $H$ gives rise to a dynamical system on $T^{*}Q$, described by Hamilton equations. These equations are the equations of motion generated by the Hamiltonian vector field $X_H\in T(T^{*}Q)$ associated with $H$. % as a solution to the equation $i_{X_H}\Omega_c=\hbox{d}H$. 
 Hamilton equations are locally described by $X_H(q,p)=\left(\frac{\partial H}{\partial p},-\frac{\partial H}{\partial q}\right)$.  that is,
\begin{equation}\label{hameq1}\dot{q}^{A}=\frac{\partial H}{\partial p_A},\quad\dot{p}_{A}=-\frac{\partial H}{\partial q^A},\quad 1\leq A\leq n.\end{equation} Equations \eqref{hameq1} determine a set of $2n$ first order ordinary differential equations (see \cite{Holmbook}, for instance, for more details).

A \textit{discrete Lagrangian} is a differentiable function
$L_d\colon Q \times Q\to \R$, which may be considered as an
approximation of the integral action defined by a continuous regular 
Lagrangian $L\colon TQ\to \R.$ That is, given a time step $h>0$
small enough,
\[
L_d(q_0, q_1)\approx \int^h_0 L(q(t), \dot{q}(t))\; dt,
\]
where $q(t)$ is the unique solution to the Euler-Lagrange equations
for $L$ with  boundary conditions $q(0)=q_0$, $q(h)=q_1$.

Construct the grid $\mathcal{T}=\{t_{k}=kh\mid k=0,\ldots,N\},$ with $Nh=T$
and define the discrete path space as 
$\mathcal{P}_{d}(Q):=\{q_{d}:\{t_{k}\}_{k=0}^{N}\to Q\}.$ We
identify a discrete trajectory $q_{d}\in\mathcal{P}_{d}(Q)$ as $q_{d}=\{q_{k}\}_{k=0}^{N}$, where $q_{k}:=q_{d}(t_{k})$. The
discrete action $\mathcal{S}_{d}:\mathcal{P}_{d}(Q)\to\R$ for this
sequence of discrete paths is calculated by summing $L_d$ on each
adjacent pair, i.e., \begin{equation*}\label{acciondiscreta}
\mathcal{S}_d(q_0,...,q_N) :=\sum_{k=0}^{N-1}L_d(q_k,q_{k+1})\approx \int^T_0 L(q(t), \dot{q}(t))\; dt.
\end{equation*}

The discrete path space is
isomorphic to the product manifold which consists of $(N+1)$
copies of $Q$. $\mathcal{S}_d$ inherits the smoothness of the
discrete Lagrangian, and the tangent space
$T_{q_{d}}\mathcal{P}_{d}(Q)$ at $q_{d}$ is the set of maps
$v_{q_{d}}:\{t_{k}\}_{k=0}^{N}\to TQ$,  with image 
$v_{q_{d}}(t_k)=\{(q_{k},v_{k})\}_{k=0}^{N}$,  such that $\tau_{Q}\circ
v_{q_{d}}=q_{d}$ where $\tau_{Q} : TQ
\rightarrow Q$ is the projection map given by $\tau_Q(q,v_q)=q$.

The discrete variational principle \cite{marsden2001discrete}, states that
the solutions of the discrete system determined by $L_d$ must
extremize the discrete action given fixed points $q_0$ and $q_N.$
Extremizing $\mathcal{S}_d$ over $q_k$ with $1\leq k\leq N-1,$ we
obtain a system of difference algebraic equations given by 
\begin{equation}\label{discreteeq}
 D_1L_d( q_k, q_{k+1})+D_2L_d( q_{k-1}, q_{k})=0,\, 1\leq k\leq N-1
\end{equation}where $D_j$ stands for the partial derivative with respect to the $j$-th component of $L_d$.

The system of algebraic difference equations \eqref{discreteeq} is known as \textit{the discrete Euler-Lagrange
equations} \cite{marsden2001discrete,hairer2006geometric}. Given a solution $\{q_{k}^{*}\}_{k\in\mathbb{N}}$ of
eq.\eqref{discreteeq} and assuming the discrete Lagrangian is regular, that is, the
matrix $(D_{12}L_d(q_k, q_{k+1}))$ is non-singular, it is possible to
define implicitly a (local) discrete flow, $
\Upsilon_{L_d}\colon\mathcal{U}_{k}\subset Q\times Q\to Q\times Q$,  by using the implicit function theorem from
(\ref{discreteeq}), as $\Upsilon_{L_d}(q_{k-1}, q_k)=(q_k, q_{k+1})$, where $\mathcal{U}_{k}$ is an open  neighborhood of the
point $(q_{k-1}^{*},q_{k}^{*})$.

\section{Lagrange-d'Alembert principle for formation control}
\label{sec: LA}
Consider a set $\mathcal{V}$ consisting of $s$ free agents evolving on a configuration manifold $Q$ with dimension $n$. We denote by $q_i\in Q$ the configurations (positions) of agent $i\in\mathcal{V}$, with local coordinates $q_i^{A}=(q_i^{1},\ldots,q_i^{n})$, and by $q=(q_1,\ldots,q_s)\in Q^{s}$ the stacked vector of positions, where $Q^{s}$ represents the cartesian product of $s$ copies of $Q$.

The neighbor relationships are described by an undirected graph $\mathcal{G}=(\mathcal{V},\mathcal{E})$ where the set $\mathcal{V}$ denotes the set of nodes, and the set $\mathcal{E}\subset\mathcal{V}\times\mathcal{V}$ denotes the set of ordered edges for
$\mathcal{G}$. The set of neighbors for agent $i$ is defined by $\mathcal{N}_i=\{j\in\mathcal{V}: (i,j)\in\mathcal{E}\}$. Since $\mathcal{G}$ is undirected, if $i\in\mathcal{N}_j$, then $j\in\mathcal{N}_i$ for the pair $(i,j)\in\mathcal{E}$.%We define the elements of the incidence matrix $B\in Q^{s\times|\mathcal{E}|}$ for $\mathcal{G}$ by 
%$$\displaystyle{
%b_{ik}=
%\begin{cases}
%+1 \hbox{ if } i=\mathcal{E}_k^{tail},\\
%-1 \hbox{ if } i=\mathcal{E}_k^{head}\\
%0 \hbox{ otherwise }
%\end{cases}
%\label{eq: B}
%}$$ where $\mathcal{E}_{k}^{tail}$ and $\mathcal{E}_k^{head}$ denote the tail and head nodes, respectively, of the edge $\mathcal{E}_k$, i.e., $\mathcal{E}_k=(\mathcal{E}_k^{tail},\mathcal{E}_k^{head})$.

The dynamics of each agent is determined by a Lagrangian system on $TQ$, that is, the motion of the agent $i\in\mathcal{V}$ is described by the Lagrangian function $L_i:TQ\to\mathbb{R}$ and its dynamics is given by the Euler-Lagrange equations for $L_i$, i.e., $$\frac{d}{dt}\left(\frac{\partial L_i}{\partial\dot{q}_i^A}\right)-\frac{\partial L_i}{\partial q_i^A}=0,\hbox{ with } i\in\mathcal{V} \hbox{ and } 1\leq A\leq n.$$

In addition, the agent $i\in\mathcal{V}$  may be influenced by a non-conservative force (conservative forces maybe included into the potential energy of each agent), which is a fibered map $F_i:TQ\to T^{*}Q$. For instance, $F_i$ can describe a virtual linear damping between two agents. 
At a given position and velocity, the force will act against variations of the position (i.e., virtual displacements). Lagrange-d'Alembert principle (or principle of virtual work) establishes that the natural motions of the forced system are those paths $q:[0,T]\to T^{*}Q$ satisfying 
\begin{equation}\label{actionforced1}\delta\int_{0}^{T}L_i(q_i,\dot{q}_i)\,dt-\int_{0}^{T}F_i(q_i,\dot{q}_i)\delta q_i\,dt=0\end{equation} for  variations vanishing at the boundary, that is, $\delta q_i(0)=\delta q_i(T)=0$ for each $i\in\mathcal{V}$.  The  first  term in \eqref{actionforced1} is  the integral action, while the second term is known as virtual work since $F_i(q_i,\dot{q}_i)\delta q_i$ is the virtual work done by the force field $F_i$ with a virtual displacement $\delta q_i$. Lagrange-d'Alembert principle leads to the \textit{forced Euler-Lagrange equations} $$\frac{d}{dt}\left(\frac{\partial L_i}{\partial\dot{q}_i^A}\right)-\frac{\partial L_i}{\partial q_i^A}=F_i(q_i^A,\dot{q}_i^A).$$ 

If  the  Lagrangian $L_i:TQ\to\mathbb{R}$ is regular,  it  induces  a  well  defined flow map,  the \textit{Lagrangian flow}, $F_t:TQ\to TQ$ given by $F_t(q_{0i},\dot{q}_{0i}):=(q_i(t),\dot{q}_i(t))$ where $q_i\in C^2([0,T],Q)$ is the unique solution of the Euler-Lagrange equation with initial condition $(q_{0i},\dot{q}_{0i})\in TQ$.

 Now consider the Lagrangian $\mathbf{L}:(TQ)^{s}\to\mathbb{R}$ defined by \begin{align}\mathbf{L}(q,\dot{q})&=\sum_{i=1}^{s}K_i(\pi_i(q),\tau_i(\dot{q}))-V_i(\pi_i(q))\label{lagrangianL}\end{align} where $K_i$ and $V_i$ are the kinetic and potential energy, respectively, of each agent, $\displaystyle{(TQ)^{s}=\Pi_{i=1}^{s}TQ}$, $\pi_{i}:Q^{s}\to Q$ the projection from $Q^{s}$ over its $i^{th}$-factor and $\tau_i:(TQ)^{s}\to TQ$ the projection from $(TQ)^{s}$ over its $i^{th}$-factor, i.e., $\pi_i(q)=q_i\in Q$ and $\tau_i(q,\dot{q})=(q_i,\dot{q}_i)$, $(q,\dot{q})\in (TQ)^{s}$.%, $q=(q_1,\ldots, q_s)\in Q^s$, $(q,\dot{q})\in (TQ)^{s}$. 
 
%=\sum_{i=1}^{s}L_i(\pi_i(q),\tau_i(\dot{q}))\label{lagrangianL}\\
 
To control the shape of the formation we introduce the local artificial potential functions $V_{ij}:Q\times Q\to\mathbb{R}$. Examples of local potentials between neighboring agents in formation control are
\begin{equation}
	V_{ij}(q_i,q_j)=\frac{1}{4}(||q_{ij}||^2-d_{ij}^2)^2,
	\label{eq: Vij}
\end{equation}
coming from distance-based control, and 
\begin{equation}\label{Vijp}
	V_{ij}(q_i,q_j)=\frac{1}{2}||q_{ij} - q_{ij}^*||^2,
\end{equation}
coming from displacement-based control. In these potentials, we have that $||\cdot||$ is a norm on $Q$ induced by the Riemannian metric on $Q$ (and therefore inducing a distance on $Q$), $q_{ij}$ denotes the relative position between agents $i$ and $j$, $d_{ij}$ denotes the desired distance between agents $i$ and $j$ for the edge $\mathcal{E}_k = (i,j)$, and $q_{ij}^*$ denotes the desired relative position between the two neighboring agents.
Note also that the artificial potentials  (\ref{eq: Vij})-(\ref{Vijp}) are not unique, and both can be given by other similar expressions as it was discussed by \cite{sun2016exponential}.

The Lagrangian function for the formation problem $\mathbf{L}_{F}:(TQ)^s\to\mathbb{R}$ is given by \begin{equation}\label{lagrangianLV}\mathbf{L}_F(q,\dot{q})=\mathbf{L}(q,\dot{q})+\frac{1}{2}\sum_{i=1}^{s}\sum_{j\in\mathcal{N}_i}V_{ij}(\pi_i(q),\pi_j(q)),\end{equation} where the factor $\frac{1}{2}$ in  \eqref{lagrangianLV} comes from the fact that $V_{ij}=V_{ji}$. For example, for each \textit{virtual spring} with \textit{elastic} potential (\ref{eq: Vij}) we have an agent at each of the tips of the spring.
 
If each agent $i\in\mathcal{V}$ is subject to external non-conservative forces, the dynamics for the formation problem is determined by an extension of Lagrange-d'Alembert principle for a single agent to multiple agents by considering the Lagrangian function $\mathbf{L}_F$. More precisely, consider the action functional \begin{equation}\label{acfuntional}\mathcal{A}(q)=\int_{0}^{T}\mathbf{L}_F(q,\dot{q})\,dt-\int_{0}^{T}F(q,\dot{q})\,dt,\end{equation}with $F:(TQ)^{s}\to(T^{*}Q)^{s}$ the stacked vector of external forces. Using the fact that the graph $\mathcal{G}$ is undirected and $V_{ij}=V_{ji}$, critical points of the action functional \eqref{acfuntional} for variations of $q\in Q^s$ with fixed endpoints and with a virtual displacement $\delta q$ for the force $F$ corresponds with the
forced Euler-Lagrange equations for $\mathbf{L}_F$ given by \begin{equation}\label{eqqforced}
\frac{d}{dt}\left(\frac{\partial L_i}{\partial \dot{q}_i^A}\right)-\frac{\partial L_i}{\partial q_i^A}+\sum_{j\in\mathcal{N}_i}\frac{\partial V_{ij}}{\partial q_i^A}=F_i,\, \, i\in\mathcal{V}.\end{equation}

\section{A variational integrator for formation control of autonomous agents}
\label{sec: fc}
The key idea of variational integrators is that the variational principle is discretized rather than the  equations of motion. 

As  in Section \ref{sec: div}, we discretize the state space $TQ$ as $Q\times Q$ and, for each agent $i\in\mathcal{V}$, let $L_{i}^d:Q\times Q\to\mathbb{R}$ be a discrete Lagrangian and let $F_{i,d}^{\pm}:Q\times Q\to T^{*}Q$ be discrete ``external forces", approximating the integral action and work done by $F_i$,  as %over a short interval of time with length $h=t_{k+1}-t_k>0$, being $h$ a constant time-step.
 \begin{align}
\int_{t_k}^{t_{k+1}}L_i(q_i(t),\dot{q}_i(t))\,dt\simeq& L_{i}^d(q_{k}^{i},q_{k+1}^{i}),\label{eqq1}\\
\int_{t_k}^{t_{k+1}}F_{i}(q_i(t),\dot{q}_i(t))\delta q_i\,dt\simeq&F_{i,d}^{-}(q_{k}^{i},q_{k+1}^{i})\delta q_{k}^{i} \label{eqq2}\\&+F_{i,d}^{+}(q_{k}^{i},q_{k+1}^{i})\delta q_{k+1}^{i}.\nonumber
  \end{align}
  
  Note that $F_i^{\pm}$ are not ``external forces'', physically speaking. They are in fact momentum, since $F^{\pm}_{i,d}$ are defined by a discretization of the work done by the force $F_i$. The idea behind the $\pm$ is that for a fixed $i\in\mathcal{V}$, one needs to combine the two discrete forces to give a single one-form   $F_{i,d}:Q\times Q\to T^{*}(Q\times Q)$ defined by \begin{equation*}\label{forcedeqdiscrete}F_{i,d}(q_0^i,q_1^i)(\delta q_0^i,\delta q_1^i)=F_{i,d}^{+}(q_0^i,q_1^i)\delta q_1^i+F_{i,d}^{-}(q_0^i,q_1^i)\delta q_0^i\end{equation*}
  
 It is known that, for a single agent (see \cite{marsden2001discrete} Section $4.2.1$), by deriving the discrete variational principle using \eqref{eqq1} and \eqref{eqq2}, one obtains the forced discrete Euler-Lagrange equations  
  \begin{align}\label{dfeleq}
0=&D_1L_i^d(q_k^i,q_{k+1}^i)+D_2L_i^d(q_{k-1}^i,q_k^i)\\&+F_{i,d}^{-}(q_k^i,q_{k+1}^{i})+F_{i,d}^{+}(q_{k-1}^{i},q_k^{i}),\,k=1,\ldots,N-1.\nonumber
\end{align}%for $k=1,\ldots,N-1$ and for a fixed $i\in\mathcal{V}$.

Equations \eqref{dfeleq} define the integration scheme $(q_{k-1}^i,q_k^i)\mapsto(q_k^i,q_{k+1}^i).$ By defining the discrete (post and pre) momenta
  \begin{align}
  p^{+}_{k,i}:=&D_2L_i^d(q_{k-1}^i,q_k^i)+F^{+}_{i,d}(q_{k-1}^{i},q_k^i),\, k=1,\ldots,N\label{momentum1}\\
  p^{-}_{k,i}:=&-D_1L_i^d(q_{k}^{i},q_{k+1}^{i})-F^{-}_{i,d}(q_{k}^i,q_{k+1}^i),\,k=0,\ldots,N-1,\nonumber
  \end{align} equations \eqref{dfeleq} lead to the integration scheme $(q_k^i,p_k^i)\mapsto(q_{k+1}^i,p_{k+1}^i)$, by writing \eqref{dfeleq} as  $p_{k,i}^{-}=p_{k,i}^{+}$.
  
In formation control, the space $(TQ)^{s}$ can be discretized as $(Q\times Q)^{s}$. For a constant time-step $h\in\mathbb{R}^{+}$, a path $q:[t_0, t_N]\to Q^s$ is replaced by a discrete path $q_d=\{q_k\}_{k=0}^{N}$ where $q_k=(q_k^1,\ldots,q_k^s)=q_d(t_k)=q_d(t_0+kh)$.
%Given that $C_d(Q^s)$ is isomorphic to $s(N+1)$ copies of $Q$ it can be endowed with a product manifold structure

%=C_d(\{t_k\}_{k=0}^{N},Q^s)
Let $C_d(Q^s)=\{q_d:\{t_k\}_{k=0}^{N}\to Q^s\}$ be the space of discrete paths on $Q^s$. Define the discrete action sum $\mathcal{A}_{d}:C_d(Q^s)\to\mathbb{R}$ by \begin{align}\mathcal{A}_d(q_d)=&\sum_{i=1}^{s}\left(\sum_{k=0}^{N-1}L_i^d(q_k^i,q_{k+1}^i)-F_{i,d}^{-}(q_k^i,q_{k+1}^{i})\delta q_k^i\right.\nonumber\\
&\left.\qquad\qquad-F_{i,d}^{+}(q_{k}^{i},q_{k+1}^{i})\delta q_{k+1}^i\right)\label{actionsum}\end{align} where to define $\mathcal{A}_d$ we are using that 
\begin{align}%x=&\int_{t_k}^{t_{k+1}}\sum_{i=1}^{s}L_i(q_i(t),\dot{q}_i(t))\,dt\nonumber\\
\int_{t_k}^{t_{k+1}}\mathbf{L}_F(q(t),\dot{q}(t))\,dt
\simeq&\left(\sum_{i=1}^{s}L_i^{d}(q_k^i,q_{k+1}^{i})\right.
\nonumber\\
&\left.+\frac{1}{2}\sum_{j\in\mathcal{N}_i}V_{ij}^{d}(q_k^i,q_{k+1}^{i},q_k^j,q_{k+1}^{j})\right)\nonumber\\&=:L^d
_F(q_k,q_{k+1})\label{dld}
\end{align}with $L^{d}_F:(Q\times Q)^{s}\to\mathbb{R}$,  $V_{ij}^d:(Q\times Q)^s\to\mathbb{R}$ a discretization of \eqref{eq: Vij} and where \begin{align*}
&\int_{t_k}^{t_{k+1}}F(q(t),\dot{q}(t))\delta q\,dt=\int_{t_k}^{t_{k+1}}\sum_{i=1}^{s}F_i(q_i(t),\dot{q}_i(t))\delta q_i\,dt\\
&\simeq\sum_{i=1}^{s}\left(F_{i,d}^{-}(q_{k}^{i},q_{k+1}^{i})\delta q_{k}^{i}+F_{i,d}^{+}(q_{k}^{i},q_{k+1}^{i})\delta q_{k+1}^{i}\right).
\end{align*}
\begin{proposition}\label{prop2discrete}
Let $L^d
_F:(Q\times Q)^s\to\mathbb{R}$ be the discrete Lagrangian defined in \eqref{dld}. A discrete path $q_d=\{q_k\}_{k=0}^{N}$ extremizes the discrete action $\mathcal{A}_{d}$ if for each $i\in\mathcal{V}$ it is a solution for the discrete forced Euler-Lagrange equations 
\begin{align}D_2L^d_i(q_{k-1}^i,q_k^i)+F^{+}_{i,d}(q_{k-1}^i,q_k^i)=&-D_1L^d_i(q_k^i,q_{k+1}^i)\label{eqpropdiscrete}\\&-F^{-}_{i,d}(q_k^i,q_{k+1}^i)\nonumber\end{align} for $k=1,\ldots,N-1$ and for variations $\delta q_k=(\delta q_k^{1},\ldots,\delta q_k^{s})$ satisfying $\delta q_0=\delta q_N=0$.
\end{proposition}

\textit{Proof:}
\textit{See Appendix A.}

Under the regularity condition $\det(D_{12}L_F^d(q_k, q_{k+1}))\neq 0$, equations \eqref{eqpropdiscrete} define implicitly a (local) discrete flow, $
\Upsilon_{L^d_F}\colon (Q\times Q)^{s}\to (Q\times Q)^s$, as $\Upsilon_{L^d_F}(q_{k-1}, q_k)=(q_k, q_{k+1})$ where $q_k=(q_k^1,\ldots,q_k^s)\in Q^s$.

In Section \ref{sec: energy} we will show that the proposed integrator has a bounded energy error, by finding error estimations for the rate of the energy dissipated along the motion of the agents at each step of the integration scheme. Another efficient discrete-times estimates for the continuous-time dynamics described by the Lagrangian $\mathbf{L}_{F}$ could be determined by the so-called lifting technique \cite{hem} (see also \cite{Lall2}).

\section{Hamilton equations and discrete Hamiltonian flow for formation control}\label{sec: ham}

%\subsection{Hamiltonian function for formation control}\label{sec: subham}
Consider $\mathbf{L}_F:(TQ)^{s}\to\mathbb{R}$ as given in \eqref{lagrangianLV}. From $\mathbf{L}_F$ we can determine the Hamiltonian function $H_F:(T^{*}Q)^{s}\to\mathbb{R}$ by defining the Legendre transform $\mathbb{F}\mathbf{L}_F:(TQ)^{s}\to(T^{*}Q)^{s}$. 

\begin{definition}
The Lagrangian system determined by $\mathbf{L}_F$ is said to be \textit{regular} if $\displaystyle{\det\left(\frac{\partial^{2} \mathbf{L}_F}{\partial \dot{q}_{i}\partial\dot{q}_j}\right)_{ns\times ns}\neq 0}$.
\end{definition}

If the kinetic energy of each agent is given by $K_i(q_i,\dot{q}_i)=\frac{1}{2}\dot{q}_i^{T}M_i\dot{q}_i$ with $M_i$ positive definite, then $\mathbf{L}_{F}$ is regular since $\displaystyle{\det\left(\frac{\partial^{2} \mathbf{L}_F}{\partial \dot{q}_{i}\partial\dot{q}_j}\right)_{ns\times ns}=\det(M)}$ with $M$ a positive definite block diagonal matrix with $s$ submatrices of dimensions $(n\times n)$ given by the matrices $M_i$.

 Therefore, one may define the Legendre transformation  $\mathbb{F}\mathbf{L}_F:(TQ)^{s}\to(T^{*}Q)^{s}$ as $\mathbb{F}\mathbf{L}_F(q,\dot{q})=(q,M\dot{q}):=(q,p)$, where $q\in Q^{s}$ and $p=(p^1,\ldots,p^s)\in(T^{*}Q)^s$ are the stacked vector of positions and momenta, respectively. For each $i\in\mathcal{V}$, $p^i=M_i\dot{q}_{i}=\frac{\partial L_i}{\partial\dot{q}_i}$, and denoting by $\bar{\tau}_i:(T^{*}Q)^{s}\to T^{*}Q$ the projection to the $i^{th}$-factor of $(T^{*}Q)^{s}$ and by $J_i$ the matrix $M_i^{-1}$, we may induce the Hamiltonian  $H_F:(T^{*}Q)^s\to\mathbb{R}$ as \begin{align}\label{hamlulti}&H_F(q,p)=\sum_{i=1}^{s}\langle\tau_i(\dot{q}),\bar{\tau}_i(p)\rangle-\mathbf{L}_F(\pi_i(q),\tau_{i}(\dot{q}(q,p)))\nonumber\\
%=&\sum_{i=1}^{s}J_i(p^i)^2\nonumber\\&-\sum_{i=1}^{s}\left(K_i(\pi_i(q),\tau_i(\dot{q}))-\frac{1}{2}\sum_{j\in\mathcal{N}_i}V_{ij}(\pi_i(q),\pi_j(q))\right)\\
=&\sum_{i=1}^{s}\frac{J_i(p^{i})^2}{2}+ V_i(\pi_i(q))-\frac{1}{2}\sum_{j\in\mathcal{N}_i}V_{ij}(\pi_i(q),\pi_j(q)).\end{align}

\begin{remark}
Note that here we are restricting our analysis to Hamiltonians where the kinetic energy for each agent $i\in\mathcal{V}$ is given by $K_i(q_i,\dot{q}_i)=\frac{1}{2}\dot{q}_i^{T}M_i\dot{q}_i$. Nevertheless, the results can be given by an abstract Hamiltonian with a general kinetic energy. In this paper, we focus on the ``standard'' kinetic energy since commonly the double integrator agents with this kinetic energy represent mobile robots in formation control \cite{oh2015survey}.
%but we restrict ourselves to such a situation since we are interested in mechanical systems which usually have such a type of kinetic energy.
\end{remark}

%\subsection{Forced Hamilton's equations}

For each $i\in\mathcal{V}$, the Hamiltonian vector field can be locally written as $X_{H_F}=\frac{\partial H_F}{\partial p^{i}}\frac{\partial}{\partial q_i}-\frac{\partial H_F}{\partial q_{i}}\frac{\partial}{\partial p^i}=\left(\frac{\partial H_F}{\partial p^i},-\frac{\partial H_F}{\partial q_{i}}\right)$, and it's integral curves are determined by Hamilton's equations \begin{equation}\label{Hamilton}\dot{q}_i^{A}=\frac{\partial H_F}{\partial p^{i}_A},\quad\dot{p}^{i}_A=-\frac{\partial H_F}{\partial q_i^{A}},\quad i\in\mathcal{V},\,\, 1\leq A\leq n.\end{equation}

Given the external force $F:(TQ)^{s}\to(T^{*}Q)^{s}$, the Legendre transformation also induces the \textit{Hamiltonian force} $F^{H_F}:(T^{*}Q)^{s}\to (T^{*}Q)^{s}$ given by $F^{H_F}=F\circ(\mathbb{F}\mathbf{L}_F)^{-1}.$

It is possible to modify the Hamiltonian vector field $X_{H_F}$ to obtain the forced Hamilton's equations, by studying the integral curves of the vector field $Z^{H_{F}}(q,p):=(X_{H_F}+Y)(q,p)$ where the vector field $Y$ is defined by  \begin{equation}\label{Yeq}Y(p_q)=\frac{d}{dt}\Big{|}_{t=0}(p_q+tF^{H_F}(p_q))\in (T^{*}Q)^{s},\end{equation} and where for each $i\in\mathcal{V}$, it is locally given by $$Y_i=F_i^{H_F}\left(q_i,\frac{\partial H_F}{\partial p^i}(q_i,p^i)\right)\frac{\partial}{\partial p^{i}}=F^{H_F}(q_i,p^i)\frac{\partial}{\partial p^i}.$$

Denoting by $Z_{H^{F}}^{i}$ the $i^{th}$-component of $Z_{H^{F}}$, \begin{align*}Z_{H^{F}}^{i}(q_{i},p^{i})=\left(\frac{\partial H_F}{\partial p^i},-\frac{\partial H_F}{\partial q_i}+F_i^{H_F}\right),\end{align*} and therefore forced Hamilton's equations are given by \begin{equation}\label{HamiltonForced}\dot{q}_i^{A}=\frac{\partial H_F}{\partial p^{i}_A},\,\dot{p}^{i}_A=-\frac{\partial H_F}{\partial q_i^{A}}+F_i^{H_F},\, i\in\mathcal{V},\, 1\leq A\leq n.\end{equation}

Using that $\displaystyle{
\frac{\partial H_F}{\partial p^{i}}=J_ip^{i}\hbox{ and }
\frac{\partial H_F}{\partial q_{i}}=\frac{\partial V_i}{\partial q_i}-\sum_{j\in\mathcal{N}_i}\frac{\partial V_{ij}}{\partial q_i}}$, forced Hamilton equations for the formation problem are
\begin{equation}\label{eqqhammulti}
\dot{q}_{i}=J_ip^i,\quad
\dot{p}^{i}=-\frac{\partial V_i}{\partial q_i}+\sum_{j\in\mathcal{N}_i}\frac{\partial V_{ij}}{\partial q_i}+F_i^{H_F},\quad i\in\mathcal{V}.
\end{equation}

Since the Hamiltonian system determined by \eqref{hamlulti} is influenced by a linear damping external forces $F^{H_F}$, the energy of the system is not preserved. The evolution of the Hamiltonian along solution curves is \begin{equation}\label{energydecay}
\frac{d}{dt}H_F(q(t),p(t))=\sum_{i=1}^{s}J_ip^{i}F_i^{H_F}(q_i,p^{i})\leq 0,
\end{equation} where the equality is given by using the solutions arising from  forced Hamilton's equations \eqref{eqqhammulti}, and the inequality is determined by using that $F_i^{H_F}=-\kappa J_ip^i$ with $\kappa>0$. Therefore the rate of change of energy decay along solutions in $(T^{*}Q)^{s}$ is determined by \eqref{energydecay}.

%\subsection{Forced Hamiltonian discrete flow}\label{sec: fl}
Given a discrete Lagrangian $L^d_F:(Q\times Q)^s\to\mathbb{R}$, the \textit{discrete Legendre transformations} $\mathbb{F}_{L^d_F}^{F^{\pm}}:(Q\times Q)^{s}\to(T^{*}Q)^{s}$ are defined through the momentum equations \eqref{momentum1} as  \begin{small}\begin{align}
\mathbb{F}_{L^d_F}^{F^{+}}(q_0,q_1)=&(q_1,D_2L^d_F(q_0,q_1)+F_{d}^{+}(q_0,q_1))=(q_1,p_1)\label{fld1}\\
\mathbb{F}_{L^d_F}^{F^{-}}(q_0,q_1)=&(q_0,-D_1L^d_F(q_0,q_1)-F_{d}^{+}(q_0,q_1))=(q_0,p_0)\label{fld2}
\end{align}\end{small}where $q_i=(q_i^1,\ldots,q_i^{s})$ and  $p_i=(p_i^1,\ldots,p_i^{s})$. %$q_1=(q_1^1,\ldots,q_1^{s})$ and  $p_1=(p_1^1,\ldots,p_1^{s})$.

If both discrete Legendre transformations are locally diffeomorphisms for
nearby $q_0$ and $q_1$, then we say that $L^d_F$ is
\textit{regular}.  Using $\mathbb{F}_{L^d_F}^{F^{\pm}}$, the forced discrete Euler--Lagrange
equations \eqref{eqpropdiscrete} can be written as
$\displaystyle{\mathbb{F}_{L^d_F}^{F^{-}}(q_k,q_{k+1})=\mathbb{F}_{L^d_F}^{F^{+}}(q_{k-1},q_k)}$.

Consider $\Upsilon_{L^d_F}\colon (Q\times Q)^{s}\to (Q\times Q)^s$ defined by Proposition \ref{prop2discrete}. It will be useful to note that
%\begin{align*}%&=\mathbb{F}_{L^d}^{F^{-}}(q_1,q_2)\\&
%\mathbb{F}_{L^d_F}^{F^{-}}\circ \Upsilon_{L^d_F}(q_0,q_1)
%&=\left(q_1,D_2L^d_F(q_0,q_1)+F_{d}^{+}(q_0,q_1))\right)\\
%&=\mathbb{F}_{L^d_F}^{F^{+}}(q_0,q_1),
%\end{align*} that is, 
\begin{equation}\label{relationF}\mathbb{F}_{L^d_F}^{F^{+}}=\mathbb{F}_{L^d_F}^{F^{-}}\circ \Upsilon_{L^{d}_F}.\end{equation}
%&=\left(q_1,-D_1L^d_F(q_1,q_2)-F_{d}^{+}(q_1,q_2))\right)\\
\begin{definition}\label{defhamin}
 We define the discrete forced Hamiltonian flow $ \widetilde{\Upsilon}_d^{F}:(T^{*}Q)^s\to (T^{*}Q)^s$ as \begin{equation}\label{sympint}
\widetilde{\Upsilon}_{d}^{F}=\mathbb{F}_{L^d_F}^{F^{-}}\circ\Upsilon_{L^d_F}\circ\left(\mathbb{F}_{L^d_F}^{F^{-}}\right)^{-1},\quad \widetilde{\Upsilon}_d^{F}(q_0,p_{0})=(q_1,p_{1}).
\end{equation} \end{definition}

Alternatively, it can also be defined as \begin{equation}\label{sympint+}
\widetilde{\Upsilon}_{d}^{F}=\mathbb{F}_{L^d_F}^{F^{+}}\circ\Upsilon_{L^d_F}\circ\left(\mathbb{F}_{L^d_F}^{F^{+}}\right)^{-1},\quad \widetilde{\Upsilon}_d^{F}(q_0,p_{0})=(q_1,p_{1}).\end{equation}

In analogy with \cite{marsden2001discrete} and \cite{ober2011discrete} we have the following results:

\begin{proposition}\label{Theo2}
The diagram in Figure \ref{fig:discretemaps} is commutative.
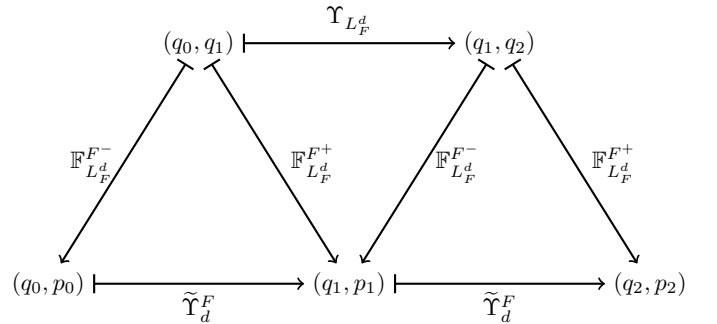
\begin{figure}[htbp]
\begin{center}
\begin{tikzpicture}[scale=0.8, every node/.style={scale=0.9}]

\path (2.5,4) node(a) {$(q_0,q_1)$}; \path
(7.5,4) node(b) {$(q_1,q_2)$}; \path (0,0)
node(c) {$(q_0,p_0)$}; \path (5,0) node(d)
{$(q_1,p_1)$}; \path (10,0) node(e)
{$(q_2,p_2)$};

\path (5.0,4.0) node[anchor=south] (f) {$\Upsilon_{L^d_F}$}; \path (1.25,2)
node[anchor=east] (g) {$\mathbb{F}_{L^d_F}^{F^{-}}$}; \path (3.9,2)
node[anchor=west] (h) {$\mathbb{F}_{L^d_F}^{F^{+}}$}; \path (6.3,2)
node[anchor=west] (i) {$\mathbb{F}_{L^d_F}^{F^{-}}$}; \path (8.9,2)
node[anchor=west] (j) {$\mathbb{F}_{L^d_F}^{F^{+}}$}; \path (2.5,0)
node[anchor=north] (k) {$\widetilde{\Upsilon}_{d}^{F}$}; \path (7.5,0)
node[anchor=north] (l) {$\widetilde{\Upsilon}_{d}^{F}$}; \draw[thick,black,|->]
(a) -- (b); \draw[thick,black,|->] (a)  -- (c);
\draw[thick,black,|->] (a)  -- (d); \draw[thick,black,|->] (b) --
(d); \draw[thick,black,|->] (b)  -- (e); \draw[thick,black,|->] (c)
-- (d); \draw[thick,black,|->] (d)  -- (e);

\end{tikzpicture}
\caption[Correspondence between the discrete Lagrangian and the
 discrete Hamiltonian map]{Correspondence between the discrete
Lagrangian and the discrete Hamiltonian maps.}\label{diagram:LH}
\label{fig:discretemaps}
\end{center}
\end{figure}

\end{proposition}
\textit{Proof: See Appendix A}
%The central triangle is \eqref{relationF}.
%The
%parallelogram on the left-hand side is commutative by \eqref{sympint}, so the triangle on the left is commutative. The triangle on the right is the same as the triangle on the left, with shifted indices. Then parallelogram on the right-hand side is commutative and therefore the triangle on the right-hand side.\hfill$\diamond$

\begin{corollary}\label{corollaryrelations} The following definitions of the discrete
Hamiltonian map are equivalent:
$
\widetilde{\Upsilon}_{d}^F=\mathbb{F}_{L^d_F}^{F^{+}}\circ \Upsilon_{L^d_F}\circ(\mathbb{F}_{L^d_F}^{F^{+}})^{-1}$,\,
$\widetilde{\Upsilon}_{d}^F=\mathbb{F}_{L^d_F}^{F^{-}}\circ \Upsilon_{L^d_F}\circ(\mathbb{F}_{L^d_F}^{F^{-}})^{-1}$,\,
$\widetilde{\Upsilon}_{d}^F=\mathbb{F}_{L^d_F}^{F^{+}}\circ(\mathbb{F}_{L^d_F}^{F^{-}})^{-1}$.\end{corollary}

\section{Discrete energy error}\label{sec: energy}\label{Sec6}

The discrete energy function associated with the formation control problem is just the discretization of the Hamiltonian $H_F$. From this observation, we propose to study the rate of energy dissipated along the motion of the agents from a Hamiltonian perspective. In particular, we will show the discrete force Hamiltonian flow $\widetilde{\Upsilon}_d$ defined in \eqref{sympint} has an asymptotically correct dissipation behavior by studying the rate of decay of a truncated modified Hamiltonian function following the approach of \textit{Backward Error Analysis} \cite{hairer2006geometric} (Chapter IX), \cite{hansen} (Sec. $4$). See also \cite{marsden2001discrete}, \cite{MdDSA}, \cite{modin}, \cite{ober2011discrete}.

 Consider the forced Hamiltonian equation \begin{equation}\label{forcedvf}\dot{x}=X_{H_F}(x)+Y(x)\end{equation} with $Y(x)=F\left(q,\frac{\partial H_F}{\partial q}\right)\frac{\partial}{\partial p}$ a vector field on $(T^{*}Q)^{s}$ as in \eqref{Yeq}, and $x=(q,p)\in(T^{*}Q)^{s}$. We aim to study Backward Error Analysis for forced variational integrators. The problem consists on finding a modified vector field $Z^{H_{F}}:=X_{H_F}(x)+Y(x)$ such that $\exp(hZ^{H_{F}})=\widetilde{\Upsilon}_{d}^F$, with  $\widetilde{\Upsilon}_{d}^F:(T^{*}Q)^{s}\to(T^{*}Q)^{s}$ being the integrator defined in \eqref{sympint} for the forced Hamiltonian system introduced in Definition \ref{defhamin}. 
 
 Since we can not invert $\exp(hZ^{H_{F}})$ to find $Z^{H_{F}}$ because the exponential map is not surjective, we must assume that $T^{*}Q$ (and hence $(T^{*}Q)^{s}$) carries a real analytic structure. Therefore, the modified vector field $Z^{H_{F}}$ can be written as an asymptotic expansion in terms of the step-size $h>0$ as \begin{equation}\label{asymptotic}Z^{H_{F}}=\sum_{k=0}^{\infty}h^{k}X_k,\end{equation}where each $X_k$ is a real analytic vector field on $(T^{*}Q)^{s}$ and it may be determined by the integrator $\widetilde{\Upsilon}_d^{F}$ for $Z^{H_{F}}$ as \begin{equation}\label{limit}X_k(q,p)=\lim_{h\to 0}\frac{\widetilde{\Upsilon}_{d}^{F}(q,p)-\exp(hX^{H_{F}}_{h,k-1})(q,p)}{h^{k}}\end{equation} with  $X_0=Z^{H_{F}}$ and $\displaystyle{X^{H_{F}}_{h,k}:=\sum_{j=0}^{k}h^{j}X_j}$.

\begin{remark}\label{remarkxk}
Note that in the construction given in \eqref{limit}, by using Taylor's theorem, it follows that $\widetilde{\Upsilon}_d^{F}(q,p)-\exp(hX_{h,k}^{H_{F}})(q,p)=\mathcal{O}(h^{k+1})$ and, if the integrator has an order $r$ then the first $r$ vector fields $X_k$ are zero.\hfill$\diamond$.
\end{remark}
 \begin{lemma}\label{lemmalip}[\cite{hairer2006geometric}, Section IX.8]
There exists a global $h$-independent Lipschitz constant for the truncated Hamiltonian $\overline{H}_F(x)=H_F(x)+\displaystyle{\sum_{k=r}^{\tau}h^{k}H_k(x)},\,\tau\in\mathbb{Z},\,x\in(T^{*}Q)^{s}$.
\end{lemma}

 The asymptotic expansion \eqref{asymptotic} does not converge in general, so, we want to find  the optimal truncation index $\tau\in\mathbb{Z}$ such that $\widetilde{\Upsilon}_d^{F}-\exp(hX_{h,\tau}^{H_F})$ converges to zero asymptotically. More formally, we want to find an order of truncation $\tau$ for \eqref{asymptotic}  \textit{depending on $h$}, such that $\displaystyle{\hbox{d}\left(\widetilde{\Upsilon}_d^{F},\exp(hX_{h,\tau}^{H_F})\right)\leq f(h)}$ with $f:\mathbb{R}\to\mathbb{R}$  continuous and $\displaystyle{\lim_{h\to 0}f(h)=0}$ for some $h\leq\alpha$ with $\alpha>0$.  The function $\hbox{d}:(T^{*}Q)^{s}\times(T^{*}Q)^{s}\to\mathbb{R}$ is given in \cite{hansen} Theorem $4.1$, and it is determined by the Whitney embedding theorem as the restriction of the Riemannian distance to an embedded submanifold of $(\mathbb{R}^{n})^{s}$.  Note that one can only choose the optimal truncation index $\tau$ in \eqref{asymptotic} if the problem has been solved, so, it is needed to implement an appropriated optimization algorithm. By also choosing an appropriated function $f$, one can, for instance, transform the problem into a convex optimization problem and optimize the truncation index $\tau$ for \eqref{asymptotic}. In the application for double integrator agents on Euclidean spaces given in this paper, one might employ a classical convex optimization algorithm \cite{boyd}. Nevertheless, in general, depending on the manifold structure one could employ specific structure-preserving convex algorithms rather than a classical one in an Euclidean space at a local level.

\begin{remark}
Note that a local Lipschitz condition is enough for mechanical systems, especially for strongly nonlinear ones, and sometimes, it is not easy to verify the global Lipschitz growth despite it always exists (see, for instance, Theorem $7.5$ in \cite{hairer2006geometric}). Moreover, for the  formation, local Lipscitz seems more appropriate. Nevertheless we maintain the original statement given in \cite{hairer2006geometric} for Lemma \ref{lemmalip}.
\end{remark}

Next, with Theorem \ref{therror} we show that the discrete force Hamiltonian flow $\widetilde{\Upsilon}_d$ has an asymptotically correct dissipation behavior, depending on the step size $h$, by studying the behavior of the discrete forced Hamiltonian flow $\widetilde{\Upsilon}_d^{F}$ for $Z^{H_{F}}$. In particular, we will show that $\overline{H}_F$ evolves with a rate of the order $\mathcal{O}(h^r)$ nearly to the exact rate of energy dissipation.

\begin{theorem}\label{therror}
Consider $\mathcal{P}:=(T^{*}Q)^{s}$ equipped with a real analytic manifold structure, $\mathcal{C}$ a compact set of $\mathcal{P}$ and $Z^{H_{F}}\in\mathfrak{X}(\mathcal{P})$ defined in \eqref{forcedvf}, real and analytic on $\mathcal{C}$. Given the discrete forced Hamiltonian flow $\widetilde{\Upsilon}_d^{F}$ for $Z^{H_{F}}$ satisfying
\begin{itemize}
\item[(1)] $\widetilde{\Upsilon}_d^{F}$ is symplectic of order $r$ when $Y=0$, 
\item[(2)] $\widetilde{\Upsilon}_d^{F}(x)$ is real and analytic for $x=(q,p)\in\mathcal{C}\subset \mathcal{P}$,
\item[(3)] there exists a sequence of real analytic vector fields $\{X_k\}_{k\in\mathbb{N}}$ on $\mathcal{P}$ with each $X_k$ as in \eqref{limit},
\end{itemize} then, there exists $\tau\in\mathbb{Z}$ depending smoothly on $h$ and positive constants $C$, $\lambda$, $\gamma$, $\alpha$, such that \begin{equation}\label{bound}
\Big{|}\overline{H}_F(\widetilde{\Upsilon}_d^{F}(x))-\overline{H}_F(x)+\Sigma(h,x,\tau)\Big{|}\leq \lambda hCe^{-\gamma/h}
\end{equation}with $h\leq\alpha$, \begin{equation}
\Sigma(h,x,\tau):=-\int_{0}^{h}\mathcal{L}_{Z^{H_{F}}_{h,\tau}}\overline{H}_F(q(s),p(s))\,ds,
\end{equation} with $(q(s),p(s))=\exp(sZ^{H_{F}})(q,p)$, and where $\overline{H}_F(q,p)$ is the truncation up to order $\tau$ of the modified Hamiltonian associated with $H_F$, that is $\displaystyle{\overline{H}_F(x)=H_F(x)+\sum_{k=r}^{\tau}h^{k}H_k(x)}$.
\end{theorem}

\textit{Proof: See Appendix A.}

\begin{remark}\label{rm: alpha}
Note that as long as the integrator evolves on the compact set $\mathcal{C}$, the Hamiltonian $\overline{H}_F$ will decreases at each step for a fixed chosen step size $h\leq\alpha$. Therefore, the rate of dissipation in energy for the discrete forced Hamiltonian flow is sufficiently close up to an order $\mathcal{O}(h^r)$ to \eqref{energydecay} for all values $x\in\mathcal{C}$ satisfying $$\sup_{x\in\mathcal{C}}|\mathcal{L}_{Z^{H_{F}}_{h,\tau}}\overline{H}_F(x)-(-JpF^{H_F}(x))|<<-JpF^{H_F}(x),$$ where the symbol $<<$ represents the magnitude order. 

%Note that $\alpha$ only appears in the last step of the proof where we needed to employ Theorem $4.1$ in \cite{hansen}.
The value $\alpha$ is crucial to get accurate simulations results and to study the convergence to the desired shape in formation control (in case there is more than one equilibrium shape). It also provides a bound on the step-size $h$ for long-time correct energy behaviors for the motion of the agents.  Note also that a time-step $h$ bigger than $\alpha$ does not guarantee the dissipation of energy of the system . We remind that the desired shape corresponds to the minimum of energy, which (in formation control) is the sum of all the energies stored by neighboring agents. %It is related with the chosen step size $h$ in order that the integrator, on points belonging to one of the coverings open sets which defines the compact set $\mathcal{C}$, evolves along a given chart on the manifold, preserving the manifold structure in the 
In the next section we will study in an application that when we work on an Euclidean space, we may use the corresponding $\alpha$ given in Theorem $8.1$ (Section IX.8) in \cite{hairer2006geometric}. Such value $\alpha$ permits to get accurate results for long-time correct energy behaviors. It must satisfy $kh\leq e^{\alpha/2h}$ with $k$ being the number of steps in the iteration of the discrete forced Hamiltonian flow and $h>0$, the time step.
\end{remark}

\section{Application of the Variational Integration in Formation Control}\label{sec: star}
\subsection{Derivation of the discretized equations of motion}
We first show how to derive the discretized equations to simulate the control of formations based on generic potentials with our proposed variational integrators.
Consider $s$ agents evolving on $Q=\mathbb{R}^n$, with local coordinates $q_i^A$, $1\leq A\leq n$, each one with unit mass. We set external forces with the linear damping $F_i(q_i,\dot{q}_i)=-\kappa\dot{q}_i, \kappa\in\mathbb{R}^+$.
%Denoting by $\Gamma_{ij}=\displaystyle{\sum_{A=1}^{n}(q_i^A-q_j^A)^2-d_{ij}^{2}}$ and 
Using \eqref{eqqforced}, the dynamics for the formation problem of the agents is given by the following set of second-order nonlinear equations
\begin{equation}\label{eq: dyncon}
\ddot{q}_i=-\kappa\dot{q}_i-\sum_{j\in\mathcal{N}_i}\nabla_{q_i}V_{ij}(q_i,q_j)\quad 1\leq i\leq s,
\end{equation}
where the potential $V_{ij}$ depends on chosen framework such as distance-based or displacement-based formation control.

To construct the numerical method, the velocities are discretized by finite-differences, i.e., $\displaystyle{\dot{q}_i=\frac{q_{k+1}^i-q_k^i}{h}}$ for $t\in[t_k,t_{k+1}]$. The discrete Lagrangian $L^d:\mathbb{R}^{sn}\to\mathbb{R}$ is given by setting the trapezoidal discretization for each Lagrangian $L_i:\mathbb{R}^{n}\times\mathbb{R}^{n}\to\mathbb{R}$ in the Lagrangian \eqref{lagrangianLV}. That is, 

\begin{align*}
L^d_i(q_k^i,q_{k+1}^i)=&\frac{(q_{k+1}^i-q_k^i)^2}{2h}\\&+\frac{h}{4}\sum_{j\in\mathcal{N}_i}(V_{ij}^d(q_k^{i},q_k^j)+V_{ij}^d(q_{k+1}^i,q_{k+1}^j)),\end{align*}where, $h>0$ is the fixed time step. 

%The discrete potential functions $V_{ij}^d$ \textcolor{blue}{for distance-based formation control} for the chosen discretization are $V_{ij}^d(q_{k}^i,q_{k}^j)=\frac{1}{4}((q_k^{i}-q_k^{j})^2-d_{ij}^{2})^{2}$, \textcolor{blue}{being $d_{ij}$ the desired distance between agent $i\in\mathcal{N}$ and $j\in\mathcal{N}_i$}.  \textcolor{blue}{For position based formation control, the discrete-time potential is given by $V_{ij}^d(q_{k}^i,q_{k}^j)=\frac{1}{2}||q_k^{i}-q_k^{j}-q_{ij}^{*}||$, with $q_{ij}^{*}$ the vector of desired positions between the agent $i$ and its neighbors.}

The external forces $F_i(q_i,\dot{q}_i)=-\kappa \dot{q}_i$ are discretized by using the trapezoidal discretization, \begin{align*}F_{i,d}^{\pm}(q_k^{i},q_{k+1}^{i})=&\frac{h}{2}F_{i}\left(q_k^{i},\frac{q_{k+1}^{i}-q_k^{i}}{h}\right)\\&+\frac{h}{2}F_{i}\left(q_{k+1}^{i},\frac{q_{k+1}^{i}-q_k^{i}}{h}\right),\end{align*} that is $F_{i,d}^{+}(q_{k-1}^{i},q_{k}^{i})=-\kappa (q_k^{i}-q_{k-1}^{i})$ and 
$F_{i,d}^{-}(q_{k}^{i},q_{k+1}^{i})=-\kappa (q_{k+1}^{i}-q_{k}^{i})$. Note that in the first term of the trapezoidal rule, the discretization chosen corresponds to a forward finite-difference and in the
second term to a backward finite-difference. Using that 
\begin{align}
D_1L_i^d(q_k^i,q_{k+1}^i)=&\frac{q_{k}^i-q_{k+1}^i}{h}+\frac{h}{4}\sum_{j\in\mathcal{N}_{i}}\frac{\partial V_{ij}^d}{\partial q_k^i}(q_k^i,q_k^j),\label{D1}\\
D_2L_i^d(q_{k-1}^i,q_{k}^i)=&\frac{q_{k}^i-q_{k-1}^i}{h}+\frac{h}{4}\sum_{j\in\mathcal{N}_{i}}\frac{\partial V_{ij}^d}{\partial q_k^i}(q_k^i,q_k^j),\label{D2}
%	\frac{\partial V_{ij}^d}{\partial q_k^i}=&\Gamma_{ij}^{k}(q_k^i - q_k^j), \label{partialx}
\end{align}  the forced discrete Euler Lagrange equations are given by \begin{equation}\label{eqqdiscrete}q_{k+1}^{i}=\kappa_hq_{k-1}^{i}+\frac{2}{1+kh}q_k^i-\bar{\kappa}_h\sum_{j\in\mathcal{N}_i}\nabla_{q_k^i}V_{ij}^{d}(q_k^i,q_k^j)\end{equation}with $\kappa_h=\frac{\kappa h-1}{1+\kappa h}$, $\bar{\kappa}_h=\frac{h^2}{2(1+\kappa h)}$, for $k=1,\ldots,N-1$. 

For example, in distance-based formation control, the equations \eqref{eqqdiscrete} are given by \begin{equation}\label{eqqdiscrete1}q_{k+1}^{i}=\kappa_hq_{k-1}^{i}+\frac{2}{1+kh}q_k^i-\bar{\kappa}_h\sum_{j\in\mathcal{N}_i}\Gamma_{ij}^k(q_k^i-q_k^j),\end{equation} where $\Gamma_{ij}^k=(q_k^i-q_k^j)^2-d_{ij}^{2}$. For displacement-based formation control, equations \eqref{eqqdiscrete} are given by \begin{equation}\label{eqqdiscrete1}q_{k+1}^{i}=\kappa_hq_{k-1}^{i}+\frac{2}{1+kh}q_k^i-\bar{\kappa}_h\sum_{j\in\mathcal{N}_i}(q_k^i-q_k^j-q_{ij}
^{*}).\end{equation}

Note that the previous equations are a set of $ns(N-1)$ for the $ns(N+1)$ unknowns $\{q_k^i\}_{k=0}^{N}$,   with $1\leq i\leq s$. Nevertheless  the boundary conditions on initial positions and velocities of the agents $q_0^{i}=q_i(0)$, $v_{q_0}^i=\dot{q}_i(0)$ contribute to $2ns$ extra equations that convert eqs. \eqref{eqqdiscrete} in a nonlinear root finding problem of $ns(N-1)$ equations and the same amount of unknowns. To start the algorithm we use the boundary conditions for the first two steps, that is, $q_0^i=q_i(0)$ and $q_1^i=hv_{q_0}^i+q_{0}^i=h\dot{q}_{i}(0)+q_{i}(0).$

\subsection{Discretized equation of the system's energy}
We now show how to derive the discretized iteration of the system's energy. Later, we will show an example of how to find a theoretical maximum step size such that the system's energy converges to zero in the case of a distance-based formation.

 Equations \eqref{eqqdiscrete} define the integration scheme by means of the 
discrete flow $\Upsilon_{L^{d}_F}:\mathbb{R}^{sn}\times \mathbb{R}^{sn}\rightarrow \mathbb{R}^{sn}\times \mathbb{R}^{sn}$ by
$\Upsilon_{L^{d}_F}(q_{k-1},q_{k})=(q_{k},q_{k+1})$, $q_k=(q_k^1,\ldots,q_k^s)$, or, by using the momentum equations \eqref{momentum1} for each $i$, the integration scheme can be written as $(q_k,p_k)\mapsto(q_{k+1},p_{k+1})$.

The total energy of each agent $E_i:TQ\to \mathbb{R}$ is given by $$E_i(q_i,\dot{q}_i)=\frac{1}{2}||\dot{q}_i||^2+\frac{1}{2}\sum_{j\in\mathcal{N}_i}V_{ij}(q_i,q_j).$$ Using the trapezoidal rule for $E_i$, the discrete energy function for each agent $E_i^d:\mathbb{R}^n\times\mathbb{R}^n\to\mathbb{R}$ is given by \begin{align*}E_i^d(q_k^i,q_{k+1}^i)&=\frac{1}{2h}(q_{k+1}^i-q_k^i)^2\\&+\frac{h}{4}\sum_{j\in\mathcal{N}_i}(V_{ij}^d(q_k^{i},q_k^j)+V_{ij}^d(q_{k+1}^i,q_{k+1}^j)).\end{align*}

Note that since $\displaystyle{\det\left(\frac{\partial^2\mathbf{L}_F}{\partial\dot{q}_{i}\partial\dot{q}_{j}}\right)=\det(\hbox{Id}_{ns\times ns})=1\neq 0}$, the Lagrangian $\mathbf{L}_F$ is regular, and therefore the Legendre transformation is a global diffeomorphism and it is given by $\mathbb{F}\mathbf{L}_F(q_i,\dot{q}_i)=(q_i,p^{i})$ where $p^{i}=\frac{\partial L}{\partial\dot{q}_i}=\dot{q}_i$. By using $\mathbb{F}\mathbf{L}_F$ we may induce the Hamiltonian function for the formation problem given by \eqref{hamlulti}. The external force $F:(TQ)^s\to(T^{*}Q)^{s}$ given by $F(q,\dot{q})=(q,-\kappa\dot{q})$ is also transformed into the Hamiltonian force $F^{H_F}:(T^{*}Q)^{s}\to (T^{*}Q)^{s}$ by using the Legendre transform, and given by $F^{H_F}(q,p)=-\kappa p$, since $\dot{q}=p$ (note that $F^{H_F}(q,p)=F((\mathbb{F}L)^{-1}(q,p))$). 

Forced Hamilton equations for  \eqref{hamlulti} are given by 
\begin{equation}\label{eq: dynconham}
	\dot{q}_i=p^i,\quad
	\dot{p}^i=-\kappa p^i+\sum_{j\in\mathcal{N}_i}\nabla_{q_i}V_{ij}(q_i,q_j).
	\end{equation}
Equations \eqref{D1}-\eqref{D2} define the  Legendre transformations as {\footnotesize \begin{align*}
\mathbb{F}_{L^d_F}^{F^{+}}(q_0^i,q_1^i)=&\left(q_1^i,\frac{1}{h}(q_1^i-q_{0}^{i})-\frac{h}{4}\sum_{j\in\mathcal{N}_i}\nabla_{q_1^i}V_{ij}^{d}(q_1^i, q_1^j)-\kappa (q_1^{i}-q_{0}^{i})\right)\\
\mathbb{F}_{L^d_F}^{F^{-}}(q_0^i,q_1^i)=&\left(q_0^i,\frac{1}{h}(q_{1}^i-q_0^i)+\frac{h}{4}\sum_{j\in\mathcal{N}_{i}}\nabla_{q_1^i}V_{ij}^{d}(q_1^i, q_1^j)+\kappa(q_{1}^{i}-q_0^{i})\right). 
\end{align*}}
Using the last two expressions and  $\Upsilon_{L^{d}_F}$, it follows the construction of the Hamiltonian flow $\widetilde{\Upsilon}_d^F$ by Corollary \eqref{corollaryrelations}.

\begin{remark}
In this work we focus on the application to formation control of double integrator agents, nevertheless the result developed here apply to a general unconstrained multi-agent mechanical control systems. Given $L_i$, $V_{ij}$ and $F_i$, one may construct $\mathbf{L}_F$ and discretize it, together with $F_i$, and the same discretization performance for the Lagrangian and the external forces. Next, it is possible to compute the discrete forced Euler-Lagrange equations, and under the regularity condition $\det(D_{12}L_F^d(q_k, q_{k+1}))\neq 0$, by solving for the step $(k+1)$ it can be defined the integration scheme.

 For applications to constrained systems (holonomic and non-holonomic), the variational integrator presented in this work for formation control can be extended in a non-trivial way. These applications to constrained systems will be studied in a further work by taking into account the results for a single agent provided in \cite{cortes2002geometric} and \cite{leyendecker2010discrete}.
\end{remark}

		\subsection{Variational Integrator vs Euler method}
\label{sec: sim}
Let us briefly review some concepts in distance-based control to grasp later the application of the variational integrators in our proposed numerical experiments. We define a desired configuration $q^*$ as a particular collection of fixed $q_i^*$ whose $SE(2)$-transformations define the \textit{desired shape}. 

Convergence results in distance-based control cover the local stabilization of the desired shape, and besides some analytical expressions for some particular cases of single-integrators \cite{mou2016undirected}, for double-integrator dynamics the neighborhoods or regions of attraction around $q^*$ (up to translations and rotations) are estimated numerically \cite{dika,de2018taming,suttern19,sun17double}. 

We say that two configurations $q^{1*}$ and $q^{2*}$ are \textit{congruent} if $||q^1_i - q^1_j|| = ||q^2_i - q^2_j||, i,j\in\mathcal{V}$ with $i\neq j$. Note that two configurations $q^{1*}$ and $q^{2*}$ can satisfy $||q^1_i - q^1_j|| = ||q^2_i - q^2_j||, (i,j)\in\mathcal{E}$ but might fail to be congruent, and therefore they \textit{do not describe the same shape}. We refer to the reader to the concept of \textit{rigidity} in formation-control \cite{anderson2008rigid} on how to construct desired shapes from a set of desired distances between agents. Therefore we can have multiple shapes corresponding to a minimum of potential functions (\ref{eq: Vij}) in distance-based control. Obviously, the more edges in $\mathcal{E}$, the more constrains and fewer possible shapes given a collection of desired distances $d_{ij}$ with $(i,j)\in\mathcal{E}$. However, in practical scenarios we are interested in keeping a small number of edges, so the system is far from an \textit{all-to-all} scheme.

It is of crucial importance in robotic multi-agent systems to choose those initial conditions, or initial deployment, for the robots such that the eventual shape is congruent with the desired one. As we will illustrate, for agents that start at rest, i.e., with $\dot q_i(0) = 0$, some desired shapes have \textit{narrow} or even disconnected regions of attractions. We find such regions after intensive campaigns of numerical simulations where we are assisted by the variational integrators proposed in this paper. In particular, we will be able to run accurate simulations with significant large time steps with the same computational cost of a simple Euler integrator. The guarantees on the decreasing of the total energy of the system over time, together with a \textit{well behavior} of such energy evolution, is of vital importance due to the high sensitivity of the gradient of the potentials (\ref{eq: Vij}) to the positions of the agents, specially when they are far from the desired shape.

We compare the performance of the variational integrator (\ref{eqqdiscrete}) and the Euler discretization of (\ref{eq: dyncon}) since both methods are similar in terms of computational cost per time step. Indeed, other methods like Runge-Kutta can give excellent results in terms of accuracy. However, one needs to evaluate the differential equation (\ref{eq: dyncon}) several times per discrete step depending on the desired accuracy, hence increasing the computational cost. We consider four agents whose desired shape is defined from a regular square $q^*$. We set $\kappa = 5$ for the dissipating forces and arbitrarily choose initial position but with the initial velocities of the agents equal to zero.
\begin{figure}[h]
\centering
\includegraphics[width=0.49 \columnwidth]{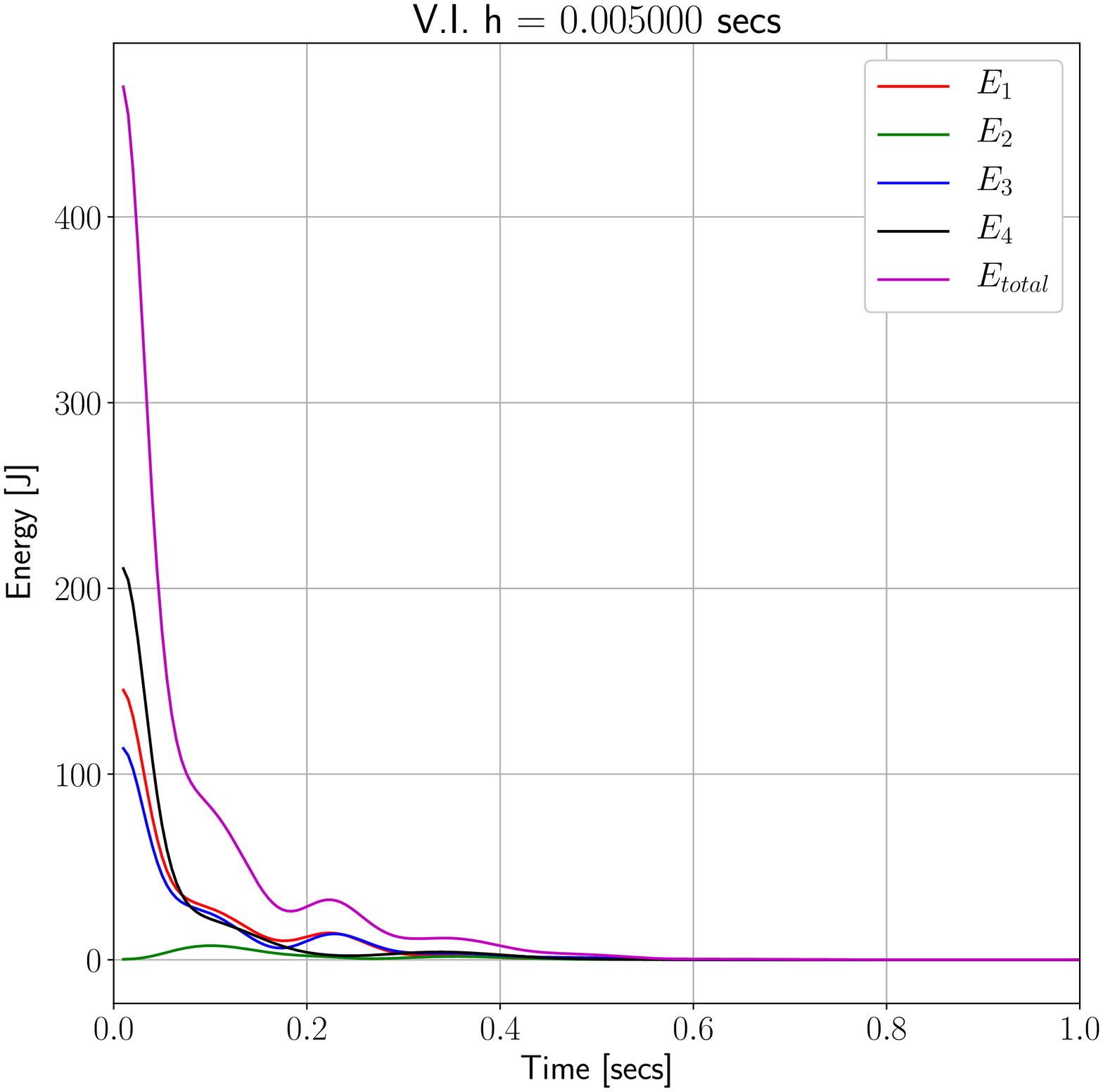}
\includegraphics[width=0.49\columnwidth]{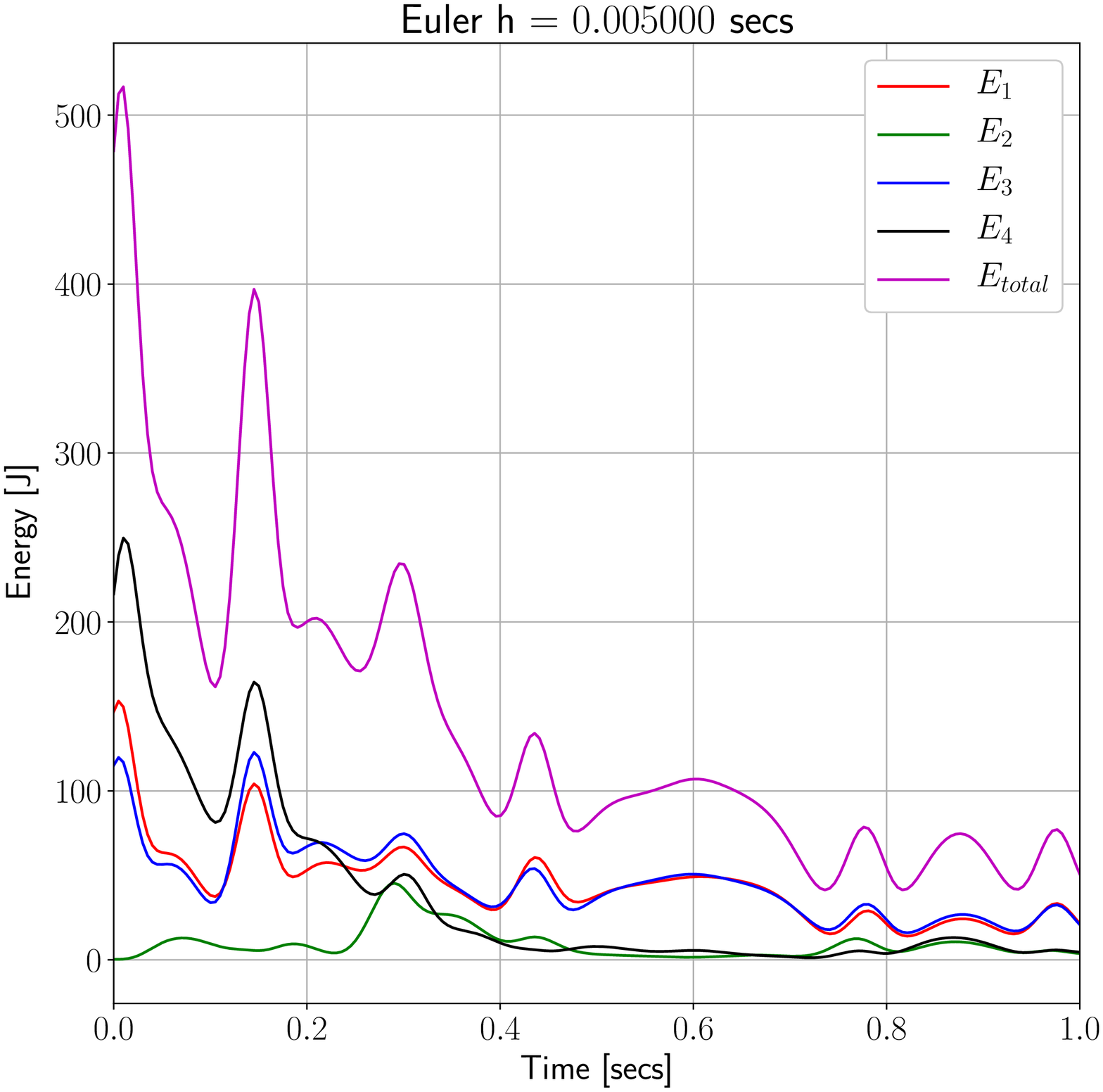}\\
\includegraphics[width=0.49\columnwidth]{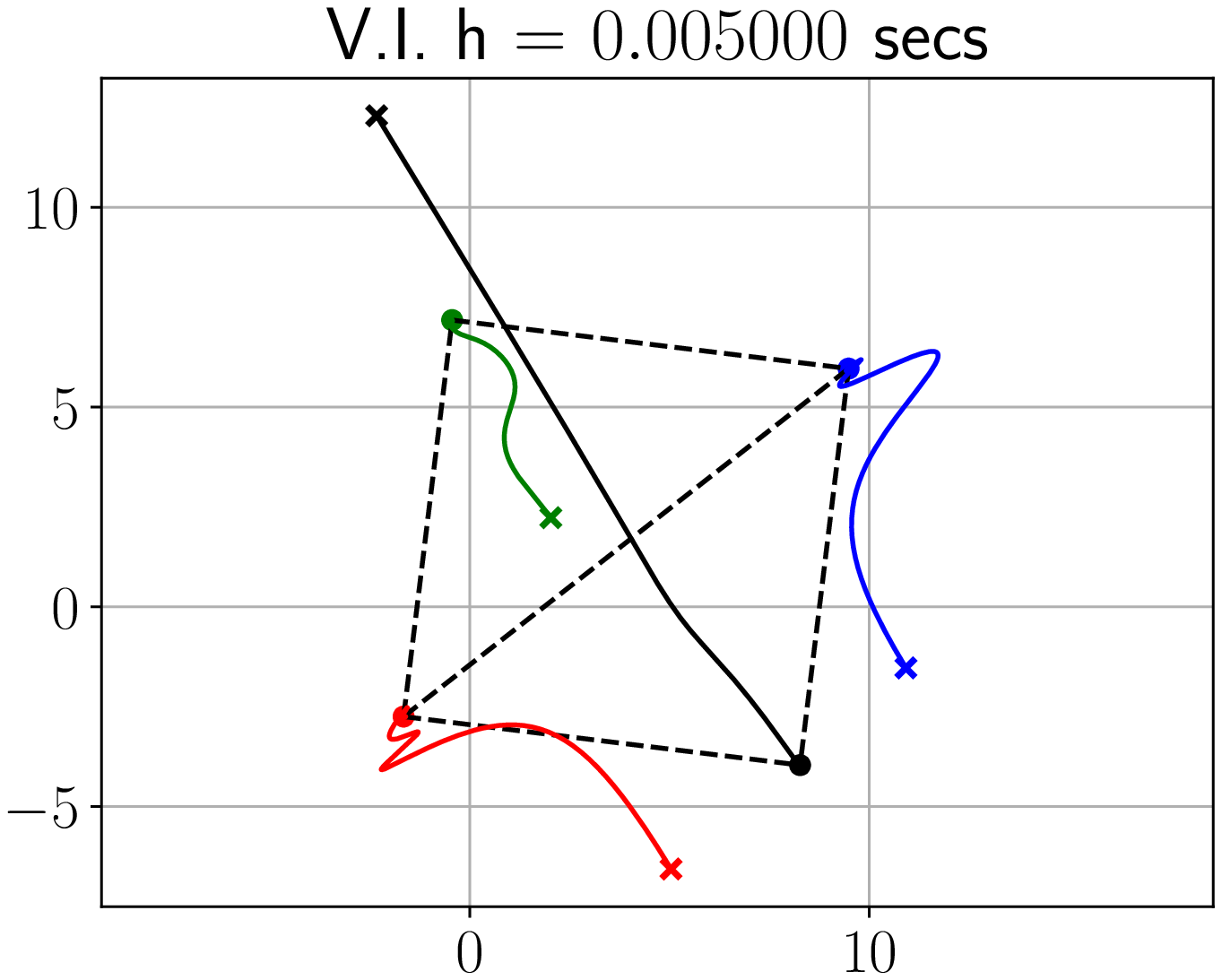}
\includegraphics[width=0.49\columnwidth]{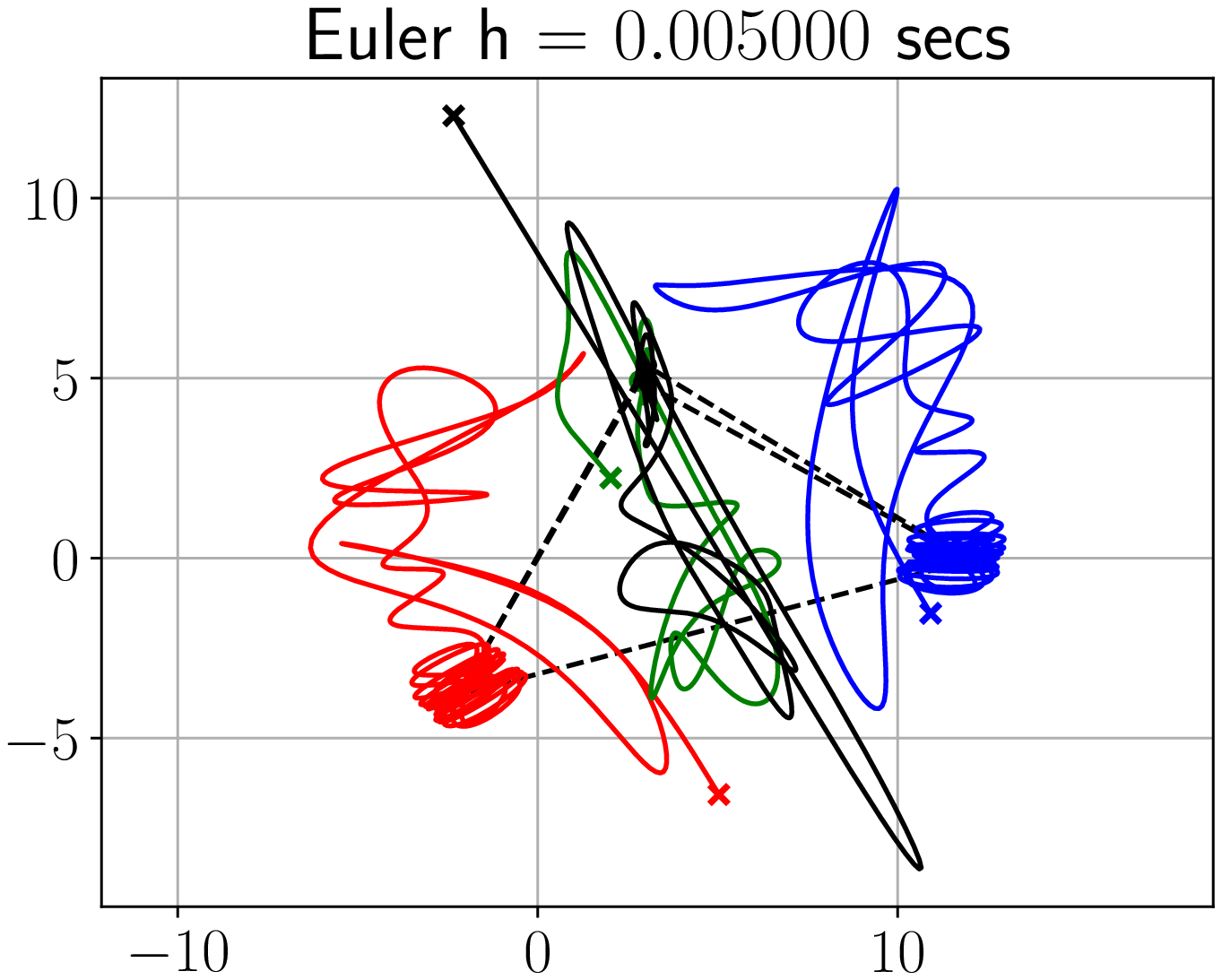}
	\caption{Comparison between the discrete energy functions of the agents, the total energy (in black color), and the agents' trajectories by employing the variational integrator (V.I.) and Euler with both having a fixed step size of $h=0.005$. The crosses denote the initial positions.}
\label{fig: sim1}
\end{figure}

\begin{figure}[h]
\centering
\includegraphics[width=0.49\columnwidth]{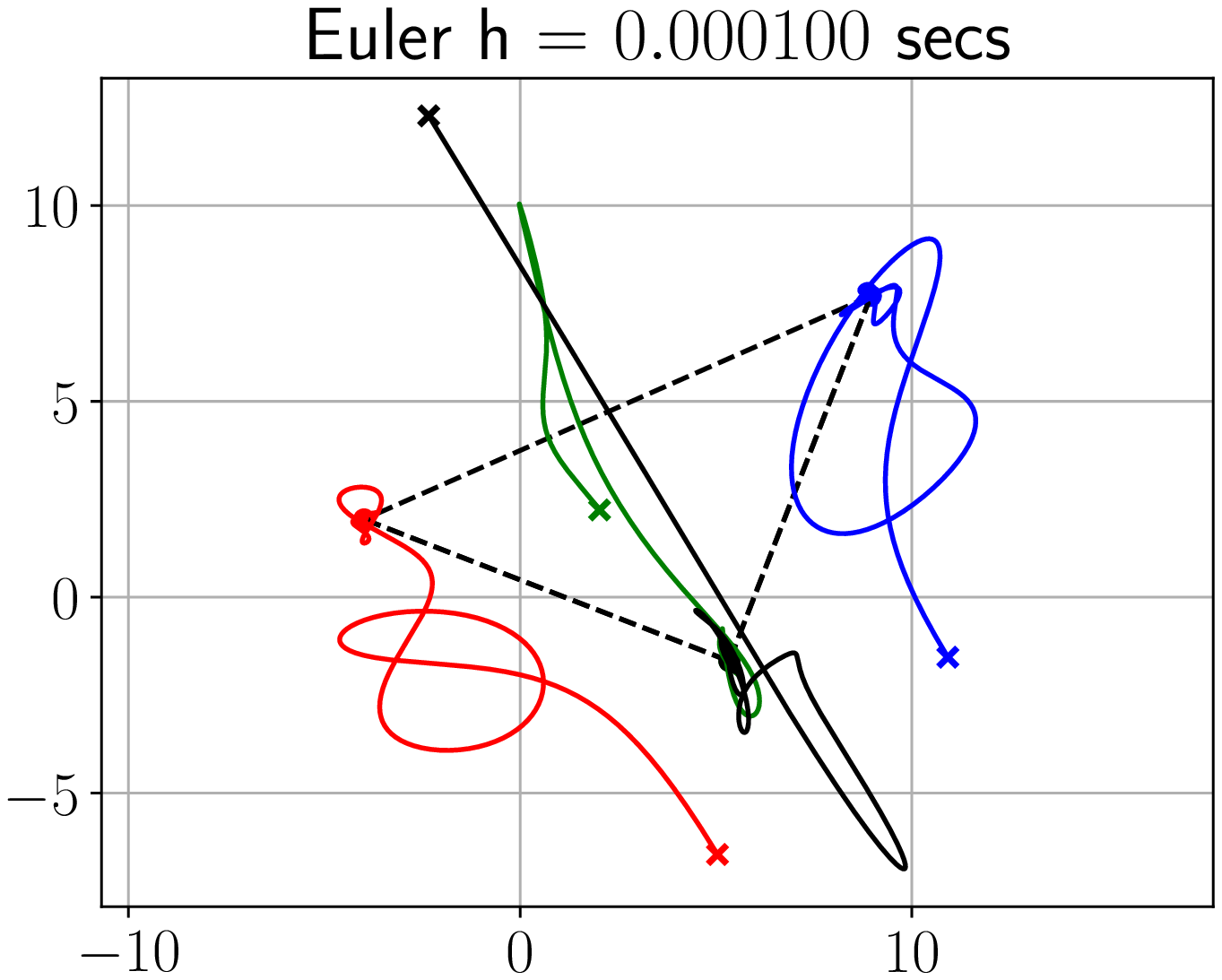}
\includegraphics[width=0.49\columnwidth]{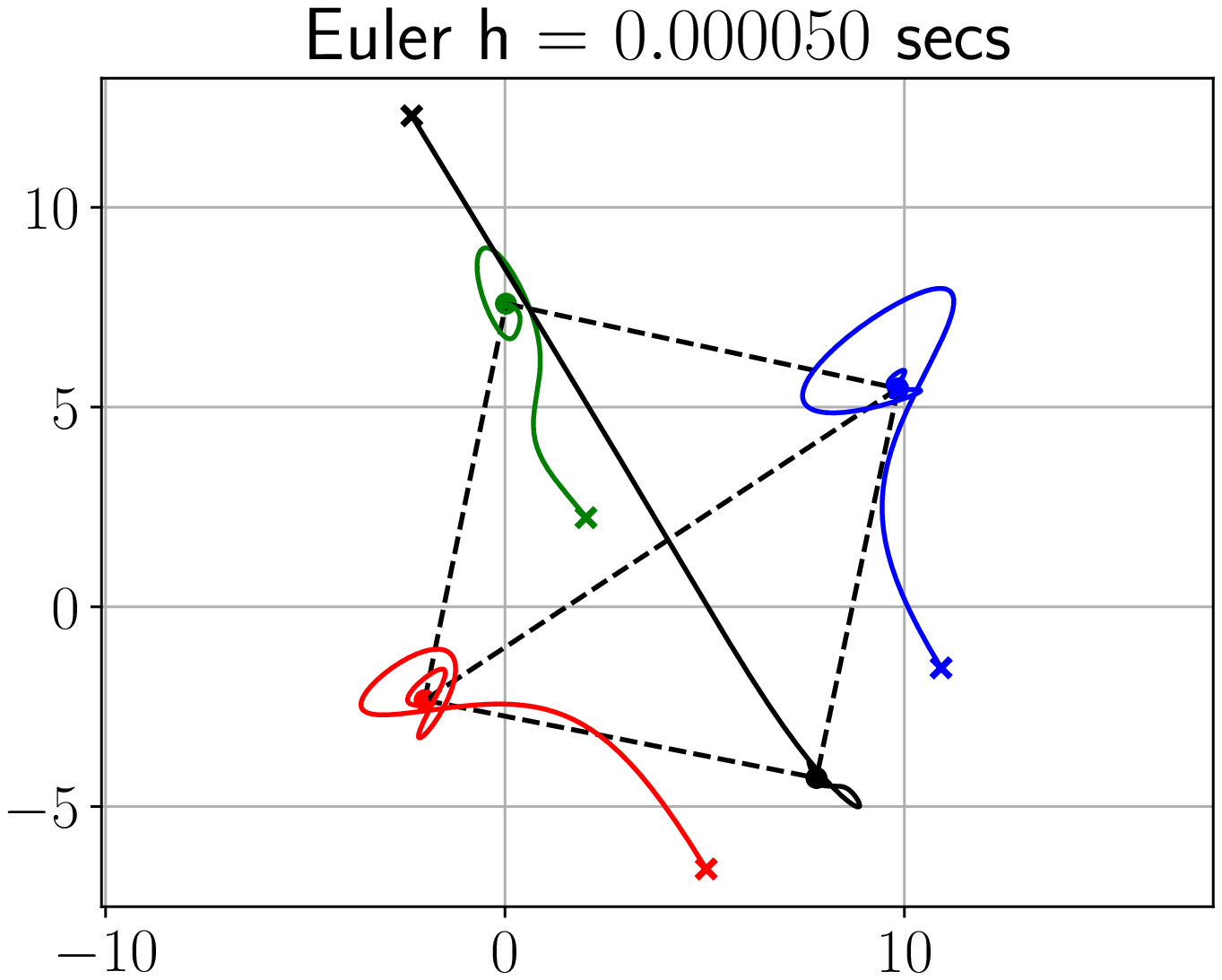}
\caption{While Euler method is stable for $h=0.05$, it is for $h=0.00005$ or lower than the transitory of the agents' trajectories, and therefore the final desired squared shape, are consistent with the results from the variational integrator. The crosses denote the initial positions.}
\label{fig: sim2}
\end{figure}

While the Euler method starts to be stable, i.e., the solution does not diverge to infinity, at $h = 0.005$, it presents a smooth behavior once the time step is lower than $h = 0.0001$. However, as it can be checked in Figures \ref{fig: sim1} and \ref{fig: sim2}, the transitory and final shapes are notably different. In fact, we only have a consistent transitory (and final desired square) when we choose $h = 0.00005$ or lower. For all the simulations we have set the number of steps to be simulated to $200$.

%That is not the case for the variational integrator, where with $h = 0.02$ we have guarantees of convergence for big set of initial conditions around the desired square $q^*$.

	\subsection{Estimation of regions of attraction in distance-based formation control}
The following numerical experiment will estimate regions of attraction for some desired shapes by exploiting the variational integration. In particular, we study the set of initial conditions for agent $i$ while the rest of agents are in the desired shape such that the eventual shape is congruent to the desired one. This case is common in practice for growing formations, and give us information on from which areas are \textit{safe} to deploy a new robot. In order to identify the region of attraction to the desired shape for one agent, we run $100$k simulations with $hr = 4$, where $r$ is the number of steps and $h$ is the time step of the variational integrator, we are also looking for those positions where the convergence time is lower than a threshold. In order to speed up the process for identifying the regions of attraction, we are interested in setting $h$ as big as possible for each simulation while having guarantees on the numerical stability, i.e., we are looking for $\alpha$ in Theorem \ref{therror}. As noted in Remark \ref{rm: alpha}, we can give the following expression for $\alpha$
\begin{equation}
	\alpha= \frac{R}{eM},\quad
	||f(q,p)|| \leq M,\quad
	||(q,p) - (q_0,p_0)||< 2R, \nonumber
\end{equation}
where $(q_0,p_0) \in \mathcal{K}:= \{(q,p) \in \mathbb{R}^{2n}\ \, \hbox{s.t.} \, ||p|| < c\}$, so for a fixed $c,R\in\mathbb{R}^{+}$ we can give a (very) conservative $M$ from (\ref{eq: dynconham}) as follows
\begin{align}
	&||f(q,p)||^2 = ||c||^2 + \sum_{i=1}^{|\mathcal{V}|}\sum_{j\in\mathcal{N}_i} ||-\kappa p_i - \nabla V_{ij}(q_{ij})||^2 \nonumber \\
	&\leq ||c||^2 + 2|\mathcal{V}|\kappa^2 ||c||^2 + 4 \sum_{(i,j)\in\mathcal{E}} ||\nabla V_{ij}(q_{ij})||^2 \nonumber \\
	&\leq (1 + 2|\mathcal{V}|\kappa^2)||c||^2\\& + 4 |\mathcal{E}| \left(\operatorname{max}_{(i,j)\in\mathcal{E}}\left\{||q|| (|\,||q_{ij}||^2 - d_{ij}^2 \,|) \right\} \right)^2 \nonumber \\
	&\leq \begin{cases}
		(1 + 2|\mathcal{V}|\kappa^2)||c||^2 + 64 |\mathcal{E}| R^6, \, \text{if} \, ||q_{ij}||^2 > d_{ij}^2, \\
				(1 + 2|\mathcal{V}|\kappa^2)||c||^2 + 64 |\mathcal{E}| R^2 \operatorname{max}\{d_{ij}^4\},\,\text{if} \, d_{ij}^2 > ||q_{ij}||^2, 
	\end{cases} \nonumber
\end{align} for $q_{ij} \in \mathcal{K},\, (i,j)\in\mathcal{E}$.

For example, in our experiment with $\kappa = 0.5$, $|\mathcal{E}| = 11$ and $|\mathcal{V}| = 7$, then for initial conditions set by $c = R = 1$ where all the agents start with $\dot p_i(0) = 0$ we have that $\alpha = 0.014$. Then, we have chosen $h = 0.014$, and with the required initial conditions, we have observed that with $200$ steps, the agents have enough time to converge to an equilibrium.

\begin{figure}
\centering
\includegraphics[width=0.49\columnwidth]{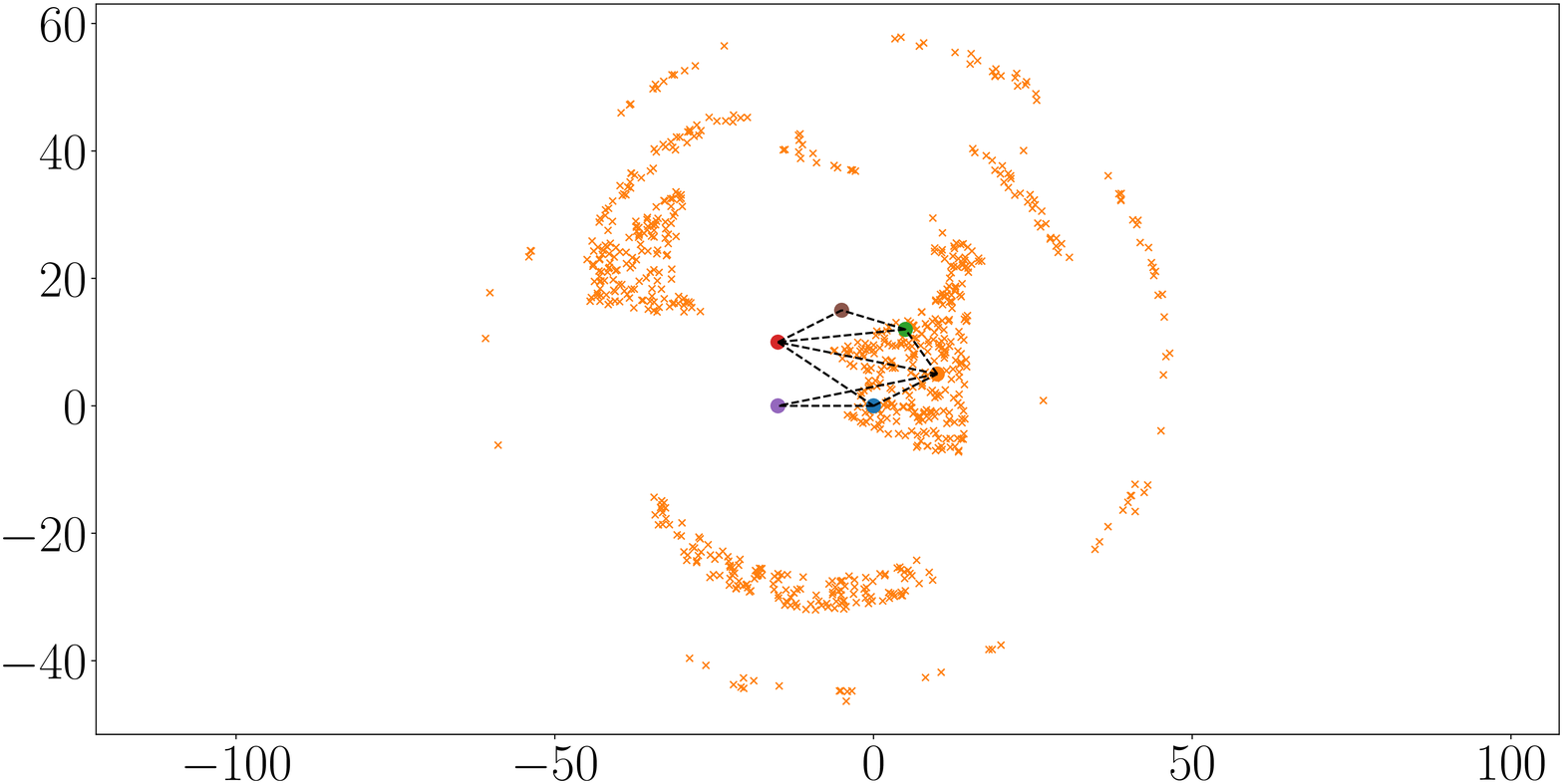}
\includegraphics[width=0.49\columnwidth]{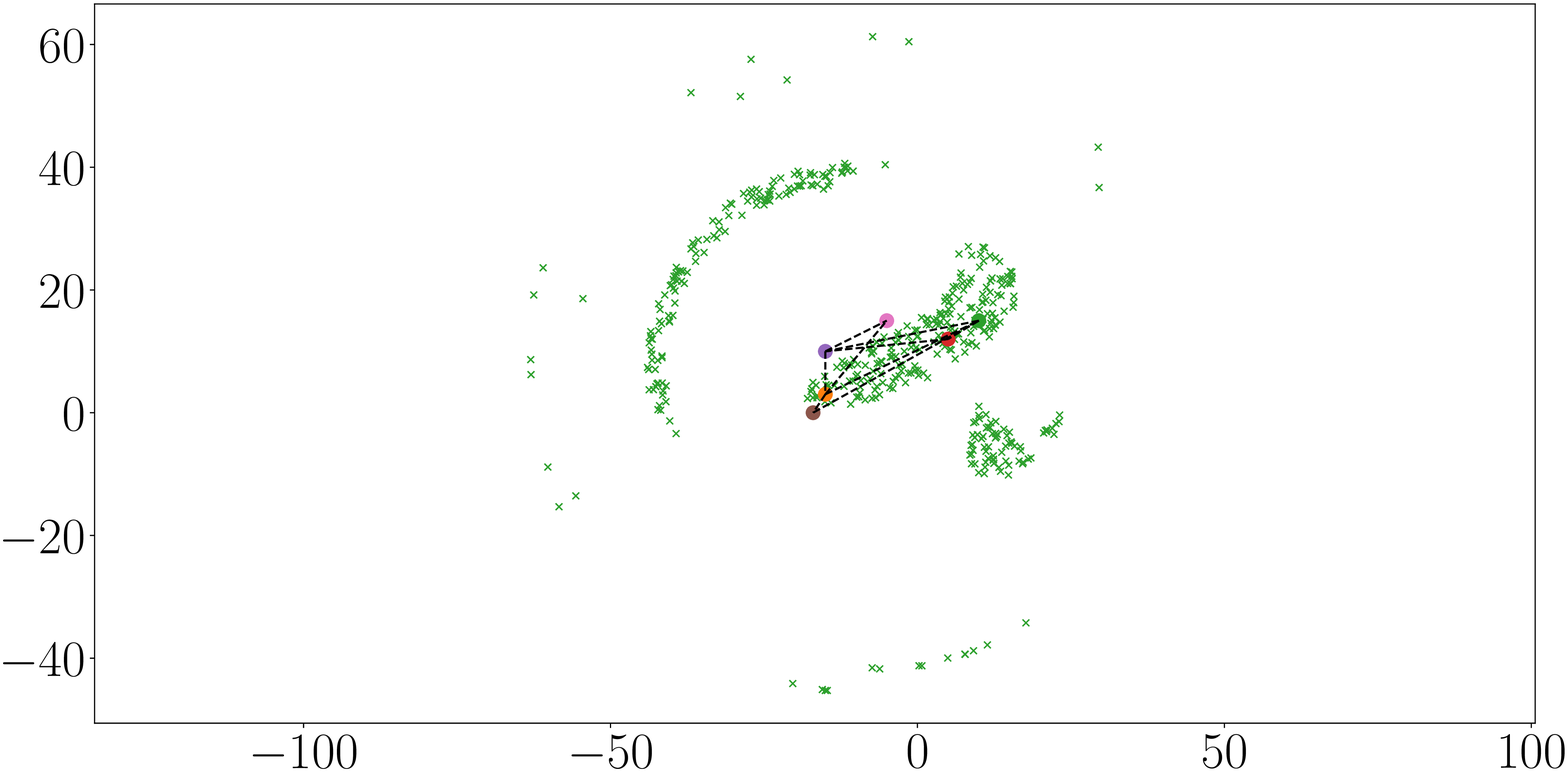}
	\caption{In these plots, all the agents except one keep all the desired distances in between at the beginning of the simulation. The Variational Integrator allows us to estimate the regions of attraction of the agent that has not been collocated correctly. This estimation is (at least four times) faster than employing Runge-Kutta 4 without losing accuracy thanks to the guaranteed well behavior of the energy of the system. Surprisingly, we identify that beyond the small perturbations of the desired position of the non-collocated agent, other areas form circular "halos" around the desired shape.}
\label{fig: sim3}
\end{figure}

To determine whether an eventual shape in a simulation is congruent to the desired one we check if the discrepancy of distances between agents in their final positions is lower than $1\%$ with respect to the desired shape in $q^*$. Indeed, we also check that the eventual velocities for the agents are also close enough to zero, e.g., $||\dot p_i(T)|| < 0.1$, being $T$ the final time of the simulation. The plots in Figure \ref{fig: sim3} show the results on regions of attraction for an arbitrary desired (rigid) shape when all the agents excepting one start at the desired shape. After testing several shapes, we estimate that for agents close to the centroid of the desired shape it is safe to start from a ball close to their desired inter-agent distances. Unexpectedly, we have identified thin halos around the centroid, but far from it, as regions of attraction, for all the tested desired shapes. More importantly, as it has been shown in the previous subsection, changing to a smaller step size does not modify the behavior of the system in the simulation as it happens with the Euler method. Therefore, one can be confident about the computed areas of attraction. Of course, the Runge-Kutta method can also give guarantees about the committed error, however, computationally is more expensive than the Variational Integrator method.

We would like to highlight that the simulation campaign with the variational integrator takes around two hours per 100k simulations in an Intel i7-2600K CPU @ 3.40GHz. 

In this simulation campaign, the integration of the equations is the most expensive operation per iteration. Therefore, the proposed variational integrator assisted us in speeding up the time-consuming process than if we were employing other methods such as Runge-Kutta 4.

\section{Conclusions and future work}
\label{sec: con}
We have presented fixed-time step variational integrators for decentralized formation control algorithms of Lagrangian systems with forcing, given that a formation problem can be seen as a Lagrangian system subject to external dissipative forces.  The paper first presented the Lagrange-d'Alembert principle for multi-agent systems in a Lagrangian mechanics framework and then we derived the forced discrete Euler-Lagrange equations from the discretization of such a variational principle. We demonstrated a general method to construct forced variational integrators for multi-agent Lagrangian, and also Hamiltonian systems. This Hamiltonian formalism allowed to formally show the rate of energy error dissipated, showing the advantage of numerical integrate the equations of motion for shape control with variational integrators compared with classical integration schemes. Consequently, we gave sufficient conditions on the step size of the numerical scheme for the stability of discrete distance-based formation control of double integrators. Finally, we have shown an application of the variational integrators assisting in a time consuming simulation campaign to identify regions of attractions of desired rigid shapes in distance-based formation control.

The methods and results given in this paper will help to numerically study and validate more complex formation control algorithms. In particular, when in practical applications we need to deal with the motion control of the formation and inconsistent measurements as it is shown in \cite{de2018taming}, or cases where a formation leader is specified, as in \cite{fl1}. In practice real-life applications, control systems are subject to perturbations and noises. In a future work, by combining the results of \cite{fl1} together with ideas of  stochastic variational integrators \cite{sto}, the proposed approach in discrete-time Lagrangian formulations and the discrete-time formation systems also extend to systems with perturbations and noises as well as flocking behavior.

\bibliographystyle{IEEEtran}
\bibliography{refs}

\section*{Appendix A}
The aim of this Appendix is to prove Theorem \ref{therror} and Propositions \ref{prop2discrete} and \ref{Theo2}. 

\textit{Proof Proposition \ref{prop2discrete}: }
Variations of the action sum \eqref{actionsum}, after a shift in the index for the discrete external force $F_{i,d}^{+}$, reads
\begin{align*}
\delta\mathcal{A}_{d}(q_d)=&\delta\sum_{k=0}^{N-1}L^d_F(q_k,q_{k+1})=D_{1}L^d_F(q_0,q_1)\delta q_0\\&+D_2L^d_F(q_{N-1},q_N)\delta q_N\\
&+\sum_{k=1}^{N-1}(D_1L^d_F(q_k,q_{k+1})+D_2L^d_F(q_{k-1},q_k))\delta q_k\\
&+\sum_{k=1}^{N-1}((F^{-}_d(q_k,q_{k+1})+F_d^{+}(q_{k-1},q_k))\delta q_k
\end{align*} where we are denoting by $\displaystyle{F^{+}_{d}(q_k,q_{k+1})=\sum_{i=1}^sF_{i,d}^{+}(q_k^i,q_{k+1}^{i})}$ and 
 $\displaystyle{F^{-}_d(q_k,q_{k+1})=\sum_{i=1}^sF_{i,d}^{-}(q_k^i,q_{k+1}^{i})}$. Requiring its stationarity for all $\{\delta q_k\}_{k=1}^{N-1}$ and $\delta q_0=\delta q_N=0$, yields the forced discrete Euler Lagrange equations 
\begin{align*}D_2L^d_i(q_{k-1}^i,q_k^i)+F^{+}_{i,d}(q_{k-1}^i,q_k^i)=&-D_1L^d_i(q_k^i,q_{k+1}^i)\\&-F^{-}_{i,d}(q_k^i,q_{k+1}^i)\end{align*} for $k=1,\ldots,N-1$ and for each $i\in\mathcal{V}$. \hfill$\diamond$

\vspace{.2cm}

\textit{Proof Proposition \ref{Theo2}:}
The central triangle is \eqref{relationF}.
The
parallelogram on the left-hand side is commutative by \eqref{sympint}, so the triangle on the left is commutative. The triangle on the right is the same as the triangle on the left, with shifted indices. Then parallelogram on the right-hand side is commutative and therefore the triangle on the right-hand side.\hfill$\diamond$

\vspace{.2cm}

\textit{Proof of Theorem \ref{therror}:} Consider the forced Hamiltonian vector field $Z^{H_{F}}(x)=X_{H_F}(x)+Y(x)$, $x\in \mathcal{P}$ given by equation \eqref{forcedvf}. From equation \eqref{energydecay} it follows that %\begin{align*}
\begin{align*}
\int_{0}^{t}-p(s)JF^{H_F}(q(s),p(s))\,ds=&H_F(\exp(tZ^{H_{F}})(x))\\&-H_F(x).\end{align*}

Consider an asymptotic expansion for $Z^{H_{F}}$, that is, $$Z^{H_{F}}_{h}=Z^{H_{F}}+\sum_{k=r}^{\infty}h^{k}(X_k+Y_k)$$ where, by Lemma \ref{lemmaH}, each $X_k$ are Hamiltonian vector fields associated with a Hamiltonian $H_k$, since $\widetilde{\Upsilon}_d^{F}$ is symplectic when $Y=0$. Note that by Remark \ref{remarkxk}, the first $r$ vector fields $X_k$ vanishes since $\widetilde{\Upsilon}_d^{F}$ is of order $r$ (i.e., from $k=0$ to $k=r-1$). We also consider a truncation order $\tau$ for $H_F$, that is, there exists $\overline{H}_F$ given by $\displaystyle{\overline{H}_F(x)=H_F(x)+\sum_{k=r}^{\tau}h^kH_k}$ with $H_k$ globally defined Hamiltonian functions associated with the vector fields $X_k$.

Denote by $Z^{H_{F}}_{h,\tau}$ the truncation of $Z^{H_{F}}_{h}$ up to an order $\tau$, that is, $\displaystyle{Z^{H_{F}}_{h,\tau}=Z^{H_{F}}+\sum_{k=r}^{\tau}h^{k}(X_k+Y_k)}$, then it follows that 

\begin{align*}
\mathcal{L}_{Z^{H_{F}}_{h,\tau}}\overline{H}_F(x)=&\langle d\overline{H}_F(x),Z^{H_{F}}_{h,\tau}\rangle\\=&\langle d\overline{H}_F(x),X_{H_F}(x)+Y(x)+\sum_{k=r}^{\tau}h^{k}(X_k+Y_k)\rangle\\
=&\langle d\overline{H}_F(x),Y(x)\rangle+ \sum_{k=r}^{\tau}h^{k}\langle d\overline{H}_F(x),Y_k\rangle
\end{align*} where we have used that $\langle d\overline{H}_F(x),X_{H_F}(x)\rangle=0$ and $\displaystyle{\sum_{k=r}^{\tau}\langle d\overline{H}_F(x),X_k(x)\rangle=0}$ since $X_{H_F}$ and $X_k's$ are Hamiltonian vector fields. Hence, given that $\langle d\overline{H}_F(x),Y(x)\rangle=-JpF^{H_F}(x)$ it follows that $$\mathcal{L}_{Z^{H_{F}}_{h,\tau}}\overline{H}_F(x)=-JpF^{H_F}(x)+\sum_{k=r}^{\tau}h^{k}\langle d\overline{H}_F(x),Y_k\rangle$$ and therefore

\begin{align}\label{eqqaa}\int_{0}^{t}\mathcal{L}_{Z^{H_{F}}_{h,\tau}}\overline{H}_F(x)(q(s),p(s))\,ds=&\overline{H}_F(\exp(tZ^{H_{F}}_{h,\tau})(x))\\&-\overline{H}_F(x)\nonumber\end{align} where $(q(s),p(s))=\exp(sZ^{H_{F}}_{h,\tau})(q,p)$.

Note that \begin{align*}\overline{H}_F(\widetilde{\Upsilon}_d^F(x))-\overline{H}_F(x)=&\overline{H}_F(\widetilde{\Upsilon}_d^{F}(x))+\overline{H}_F(\exp(tZ^{H_{F}}_{h,\tau})(x))\\&-\overline{H}_F(\exp(tZ^{H_{F}}_{h,\tau})(x))-\overline{H}_F(x),\end{align*} then using \eqref{eqqaa}, it follows that 
\begin{align}
&\Big{|}\overline{H}_F({\widetilde{\Upsilon}_d^{F}(x)})-\overline{H}_F(x)-\int_{0}^{t}\mathcal{L}_{Z^{H_{F}}_{h,\tau}}\overline{H}_F\left(\exp(sZ^{H_{F}}_{h,\tau})(x)\right)\,ds\Big{|}\nonumber\\&=\Big{|}\overline{H}_F(\widetilde{\Upsilon}_d^{F}(x))-\overline{H}_F\left(\exp(tZ^{H_{F}}_{h,\tau})(x)\right)\Big{|}\label{number1} .
\end{align} 

Finally, by using Lemma \ref{lemmalip}, there exists $\lambda>0$ such that \eqref{number1} at time $t=nh$ is upper bounded by $\lambda \hbox{d}(\widetilde{\Upsilon}_d^{F}(x), \exp(hZ^{H_{F}}_{h,\tau})(x))$ with $\hbox{d}$ the distance function given in Theorem \ref{hansen}.  Therefore, by applying again Theorem \ref{hansen} to the last expression we have that 
\begin{align*}
\Big{|}\overline{H}_F(\widetilde{\Upsilon}_d^{F}(x))&-\overline{H}_F\left(\exp(tZ^{H_{F}}_{h,\tau})(x)\right)\Big{|}\\
&\leq \lambda \hbox{d}(\widetilde{\Upsilon}_d^{F}(x), \exp(hZ^{H_{F}}_{h,\tau})(x))\leq \lambda Che^{-\gamma/h}
\end{align*} for some $h\leq\alpha$ with $\alpha>0$ small enough.\hfill$\diamond$

\section*{Appendix B}

The aim of this Appendix is to provide the basic definitions about geometric integration we used to prove Theorem \ref{therror}.

%	\subsection{\textcolor{blue}{Review on geometric integration }}\label{subsec}
%\textcolor{blue}{Next, we introduce the basic definitions on the geometry of Hamiltonian systems we will use along Sections \ref{sec: ham} and \ref{Sec6}. The reader can skip this subsection and \ref{subsec} if the interest is within the implementation of the variational integrator  rather than  a detailed analysis for the asymptotically behavior of the energy dissipated along the motion of the agents.}

Consider the ordinary differential equation \begin{equation}\label{ode}\frac{d}{dt}y(t)=X(y(t)),\end{equation} with $X$ a vector field on a manifold $Q$ and $y(t)\in Q$. The flow map for $X$ is denoted by $R:\mathbb{R}\times Q\to Q$. We use the notation $R_{X}(t,q)$ to specify the associated vector field or simply $R_{X,t}(q)$. The flow $R_{X,t}$ may be given by the exponential map as $R_{X,t}(q)=\exp(tX)(q)$, where $t$ is a parameter and $\exp:\mathfrak{X}(Q)\to\hbox{Diff}(Q)$, with $\hbox{Diff}(Q)$ denoting the set of diffeomorphisms on $Q$ and $\mathfrak{X}(Q)$ the set of vector field on $Q$. In the following, we assume that the flow $\exp(tX)$ is explicitly integrable, and therefore one may use a classical integrator as an Euler's method to compute the flow.

 Under this assumption, a numerical approximation to the solution of \eqref{ode} can by given by constructing a family of diffeomorphisms $\{\Phi_h\}_{h\geq 0}$ and then, for each $h$ fixed, it may be possible to obtain the sequence $\{q_{h,n}\}_{n\in\mathbb{N}}$ satisfying $\Phi_h(q_{h,n})=q_{h,n+1}$, called \textit{numerical integrator}. 

\begin{definition}
An \textit{integrator} for $X$ is a family of one-parameter diffeomorphisms $\Phi_{h}:Q\to Q$ (smooth in $h$) satisfying $\Phi_{0}(x)=x$ with $x\in Q$, and $\Phi_{h}(x)-\exp(hX)(x)=\mathcal{O}(h^{r+1})$ with $r\geq 1$ being the order of the integrator.
\end{definition}

\begin{definition}
An integrator $\Phi_h$ is called \textit{symplectic} if it is a symplectic diffeomorphism with respect to the symplectic canonical structure $\Omega_c$ on $T^{*}Q$ for each $h>0$ (see \cite{Holmbook} for instance).
\end{definition}

 \begin{lemma}\label{lemmaH}[\cite{hairer2006geometric}, Section IX.3]
If $\Phi_h$ is a symplectic integrator, then each vector field $X_k$ on \eqref{asymptotic} is a Hamiltonian vector field and therefore each of these vector fields is associated to a Hamiltonian function $H_k$.
\end{lemma}

\vspace{.1cm}

Along the proof of Theorem \ref{therror}, we will use the following result where $\Phi_h$ must be considered as $\Phi_h:=\varphi\circ\Phi_h\circ\varphi^{-1}$ for a given local chart $(U,\varphi)$ on $Q$. 
\vspace{.1cm}
\begin{theorem}\label{hansen}[A. C. Hansen (2011) Theorem $4.1$ \cite{hansen}]
Let $\mathcal{M}$ be a real and analytic smooth manifold, $\hbox{d}$ a metric on $\mathcal{M}$, $X$ a real analytic vector field on $\mathcal{M}$ and $\Phi_h$ be an integrator for $X$ of order $r$ such that $h\mapsto\Phi_h(q)$ is analytic for $q\in\mathcal{K}\subset\mathcal{M}$ with $\mathcal{K}$ compact. There exists $\tau\in\mathbb{Z}$ depending on $h$ and positive constants $C,\alpha,\gamma$ such that for $\displaystyle{X_{h,\tau}=\sum_{j=1}^{\tau}h^{j-1}X_j}$ it follows that $\displaystyle{\hbox{d}\left(\Phi_h(q),\exp(hX^{H_{F}}_{h,\tau})(q)\right)\leq Che^{-\gamma/h}}$ for all $q\in\mathcal{K}$ and $h\leq \alpha$. 
\end{theorem}

\end{document}